\documentclass[letterpaper,12pt,twoside]{article}
\usepackage{amsfonts}
\usepackage{amsmath}
\usepackage{amsopn}
\usepackage{amsthm}
\usepackage{amssymb}
\usepackage{graphicx}
\usepackage{rotating}
\usepackage[]{natbib}
\usepackage[english]{babel}
\usepackage[utf8]{inputenc}
\usepackage[colorlinks=true,citecolor=blue,linkcolor=blue]{hyperref}
\usepackage{xspace}
\usepackage{setspace}
\usepackage{dsfont}
\usepackage{booktabs}
\usepackage[T1]{fontenc}
\usepackage{anysize}
\usepackage{enumerate}
\usepackage[labelfont=bf, labelsep=period]{caption}
\marginsize{1.87cm}{1.87cm}{1.87cm}{1.87cm} % left, right, top , bottom
\usepackage{fancyhdr}
\RequirePackage{txfonts}
\usepackage{times}
\usepackage{bm}
\usepackage[linkcolor=blue,colorlinks=true]{hyperref}

\newtheorem{Proposition}{Proposition}

\newtheorem{Theorem}{Theorem}
\newtheorem{Lemma}{Lemma}
\newtheorem{Assumption}{Assumption}
\newtheorem{Remark}{Remark}
\DeclareMathOperator{\re}{\mathbb{R}}
\DeclareMathOperator{\na}{\mathbb{N}}
\newcommand{\E}{\mathbb{E}}
\newcommand{\ind}{\mathds{1}}
\newcommand{\e}{\mathrm{e}}
\newcommand{\Prob}{\mathbb{P}}
\def\bx{\mathbf{x}}
\def\bX{\mathbf{X}}
\def\by{\mathbf{y}}
\def\bY{\mathbf{Y}}
\def\bZ{\mathbf{Z}}
\def\bD{\mathbf{d}}
\def\bH{\mathbf{H}}
\def\d{\mathrm{d}}
\def\bbeta{\boldsymbol\beta}
\def\bmu{\boldsymbol\mu}
\def\bSigma{\boldsymbol\Sigma}
\def\bLambda{\boldsymbol\Lambda}

\def\btheta{\boldsymbol\theta}
\def\sgn{\text{sgn}}
\newcommand{\I}{\mathbf{I}}
\def\Oset{\mathrm{O}}

\allowdisplaybreaks

\begin{document}

\def\figureautorefname{Figure}
\def\sectionautorefname{Section}
\def\subsectionautorefname{Section}
\def\subsubsectionautorefname{Section}
\def\Propositionautorefname{Proposition}
\def\Theoremautorefname{Theorem}
\def\Lemmaautorefname{Lemma}
\def\Assumptionautorefname{Assumption}
\renewcommand*\footnoterule{}

\title{Reconciling Bayesian and frequentist approaches to robustness against outliers}

\author{Philippe Gagnon$^{1}$ and Alain Desgagné$^{2}$}

\maketitle

\thispagestyle{empty}

\noindent $^{1}$Department of Mathematics and Statistics, Universit\'{e} de Montr\'{e}al.

\noindent $^{2}$Department of Mathematics, Université du Québec à Montréal.

\begin{abstract}
Heavy-tailed models are often used as a way to gain robustness against outliers in Bayesian analyses. In frequentist analyses, M-estimators are often employed. In this paper, the two approaches are tentatively reconciled by considering M-estimators as maximum likelihood estimators of heavy-tailed models. From this perspective, it is realized that a fundamental difference exists as frequentists, contrarily to Bayesians, do not require these heavy-tailed models to be proper. For instance, a popular robust estimator in linear regression, Tukey's biweight M-estimator, does not correspond to a proper heavy-tailed model. Thus, a Bayesian practitioner does not have access to the same range of tools as a frequentist practitioner. It is shown through two real-data linear regression analyses that the former may in consequence obtain significantly different estimation results than the latter, where the difference is due to a more pronounced influence by the outliers in the former case. It is highlighted that a way to give these practitioners access to the same range of tools is for the Bayesian to adopt the generalized Bayesian framework of \cite{bissiri2016general} which allows the use of improper models \citep{jewson2022general}, in combination with proper prior distributions yielding proper generalized posterior distributions. A complete reconciliation of the Bayesian and frequentist approaches to robustness is then achieved. An extensive theoretical study of the generalized Bayesian counterpart of Tukey's biweight M-estimator is provided, which includes a robustness characterization result, a strong consistency result and a Bernstein--von Mises result, the latter allowing to calibrate the generalized posterior distribution for meaningful uncertainty quantification. After adopting the generalized Bayesian framework, the Bayesian practitioner obtains similar results as the frequentist practitioner in the aforementioned examples.
\end{abstract}


\section{Introduction}\label{sec:intro}

\subsection{Context}\label{sec:context}

Let us assume that we have access to a data set of the form $\{\bx_i, y_i\}_{i=1}^n$, where $\bx_1, \ldots, \bx_n \in \re^p$ are $n \in \na$ vectors of explanatory variables and $y_1, \ldots, y_n \in \re$ are $n$ observations of a dependent variable, $p$ being a positive integer. Let us assume that one is interested in modelling the dependent variable through its relationship with the explanatory variables, and therefore in using a regression model.

It is common that the data set used for model estimation is contaminated by outliers (i.e., erroneous or extreme data points). Suppose that there exists a trend in the bulk of the data (i.e., the non-outliers), and that the model is used to capture this trend. We define an outlier as a couple $(\bx_i, y_i)$ whose components are incompatible with this trend. It is thus not necessary that either component, $\bx_i$ or $y_i$, be extreme (in the sense of being large in norm); rather, it is the combination of this $\bx_i$ with that $y_i$ that makes the couple an outlier. Note that outliers are model-dependent (because, as mentioned, the model is used to capture the trend in the bulk of the data and outliers are couples  $(\bx_i, y_i)$ whose components are incompatible with this trend). Robust approaches provide statistical conclusions that are consistent with the trend in the bulk of the data in the presence of outliers. The global objective of this work is to investigate the connection between classical Bayesian and frequentist approaches to robustness against outliers.

\subsection{Connecting Bayesian and frequentist robust approaches}\label{sec:connection}

Typical regression models are not robust against outliers, meaning that the difference in trends in the bulk of the data and the outliers results in skewed inference and predictions. A canonical example of non-robust models is a linear regression with normal errors for which maximum likelihood estimation of the regression coefficients corresponds to ordinary least squares (OLS). For this model (and in fact many others), the cause of the non-robustness is the exponential decay of the tails which makes the presence of outliers supporting a different trend unlikely and the model simply not adapted to their presence \citep{o1979outlier}. Non-robustness is thus related to the notion of model misspecification.

The classical Bayesian approach to deal with a robustness problem is to replace the light-tailed distribution by one that has heavy tails, while being similar to the light-tailed one. The resulting model increases the likelihood of outliers and is adapted to their presence. It is thus robust, while leading to inference similar to the original model in the absence of outliers, a desirable property referred to as \textit{efficiency} in a robustness context. Another desirable property is \textit{whole robustness} \citep{desgagne2015robustness}, according to which the robust approach acts automatically as practitioners would and excludes outliers when they are far enough away and there is no doubt as to whether they really are outliers. An efficient and wholly robust approach thus also leads to similar estimation results as with the original model when the outliers are far enough, but where the estimation of the original model in this case is based on the data set without the outliers (the bulk of the data). The data set without the outliers is often expected and observed in practice (as in the examples in this article) to be more closely aligned with the assumptions of the original model. A final desirable property is a gradual diminution of the impact of observations when artificially moved away from the bulk of the data. This property allows for a certain influence of moderately far observations, reflecting uncertainty about the nature of these observations in a grey zone (outliers versus non-outliers).

In Bayesian linear regression, a Student's $t$ distribution is often used as a robust alternative \citep{1984west431}. Throughout the years it has been found that, even though heavy, the tails of the Student's $t$ probability density function (PDF) are not heavy enough to reach whole robustness in linear regression \citep{gagnon2023theoretical}. The \emph{log-Pareto-tailed normal} (LPTN), a PDF introduced in \cite{desgagne2015robustness}, has sufficiently heavy tails to reach whole robustness \citep{gagnon2020}. The central part of this continuous PDF coincides with that of the standard normal and the tails are log-Pareto, meaning that they behave like $(1/|z|)(1 / \log |z|)^\lambda$ with $\lambda > 1$, hence its name. The LPTN PDF belongs to a family of PDFs introduced in \cite{desgagne2015robustness} which are referred to as \textit{log-regularly varying}.

On the frequentist side, the classical approach to gain robustness consists in the derivation of robust estimators through a modification of the log-likelihood function or its derivative. In normal linear regression, the source of the robustness problem is identified as the quadratic function applied to the residuals in the log-likelihood function which significantly increases the impact of outliers. This quadratic function is replaced by one that grows less rapidly. When the modification is regarding the log-likelihood function, the approach yields what is referred to as an \textit{M-estimator} \citep{huber1964robust}.

The rationale behind both Bayesian and frequentist approaches is the same (limiting the impact of outliers) and a connection is apparent as they both consist of a modification of the likelihood function (or its derivative). For instance, the Huber M-estimator \citep{huber1973robust} in linear regression can be seen as the maximum likelihood estimator of a model where the PDF of the errors has tails with the same decay as those of the Laplace PDF; this PDF thus has heavier tails than the normal. The frequentist approach to robustness can thus be seen as corresponding to the use of heavy-tailed distributions as well. There is, however, a fundamental difference between the frequentist and traditional Bayesian approaches to robustness. On the one hand, the modified log-likelihood function (in the context of M-estimation) is not required to correspond to a proper model. For instance, Tukey's biweight M-estimator \citep{beaton1974fitting} in linear regression does not correspond to a proper model as the modified log-likelihood function is constant beyond a threshold, thus yielding an improper distribution. On the other hand, traditional Bayesians require the model to be proper and thus do not have access to such improper distributions.

A contribution of this paper is to highlight this fundamental difference. Another contribution is to demonstrate that a practitioner may obtain significantly different estimation results depending on whether the practitioner is a frequentist or a traditional Bayesian, where the results in the latter case are more influenced by the presence of outliers, even when using the super heavy-tailed LPTN PDF. In this paper, we focus on linear regression for the explanations and examples, but it is clear that the aforementioned fundamental difference between the frequentist and traditional Bayesian approaches to robustness exists and has consequences beyond the context of linear regression.

\begin{figure}[ht]
 \centering\small
 $\begin{array}{cc}
   \includegraphics[width=0.5\textwidth]{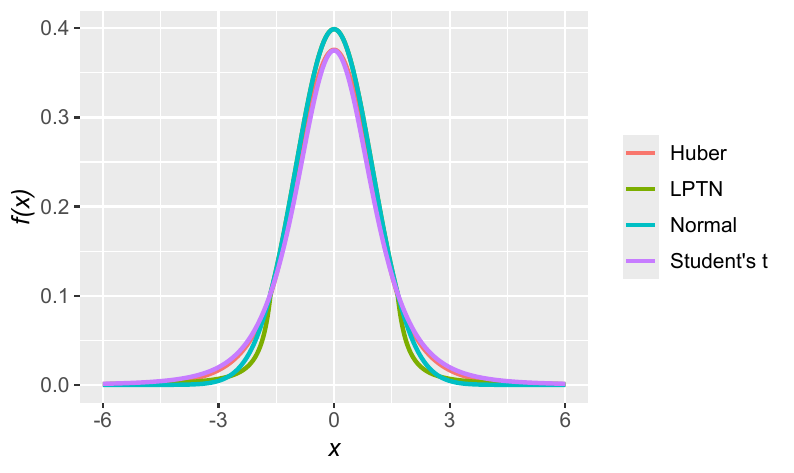} & \hspace{-5mm} \includegraphics[width=0.5\textwidth]{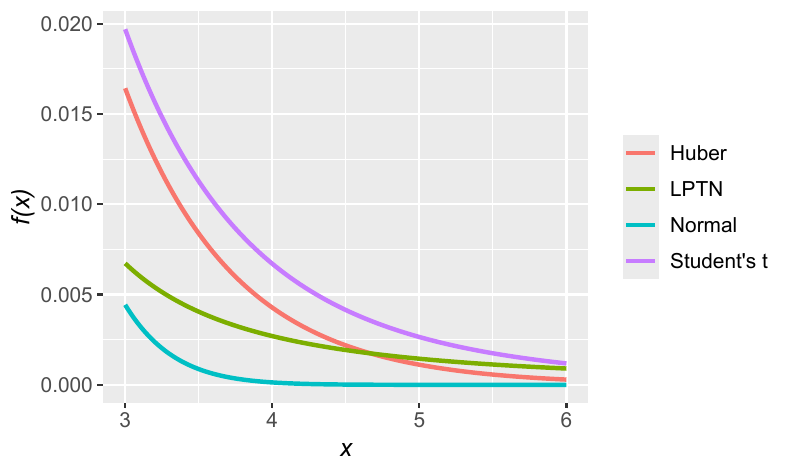} \cr
   \hspace{-10mm} \textbf{(a) } & \hspace{-14mm} \textbf{(b)}
  \end{array}$
  \vspace{-3mm}
\caption{(a) PDF associated with Huber loss function and LPTN, normal and Student's $t$ PDFs; the normal and LPTN PDFs appear similar, and the Huber and Student's $t$ PDFs appear similar. (b) Zoom on right tail. The density associated with Tukey's biweight loss function is not shown as its improperness makes difficult a direct comparison. }\label{fig:comparisons_PDFs}
\end{figure}

The classic frequentist robust approaches in the context of linear regression are arguably those mentioned above, namely the Huber and Tukey's biweight M-estimators. On the Bayesian side, as mentioned, the Student's $t$ model is the classic approach. The only other Bayesian heavy-tailed approach that has been proposed is that of using a log-regularly varying distribution \citep{gagnon2020}, with the LPTN distribution being the most popular. When the M-estimators are considered as maximum likelihood estimators of (possibly improper) heavy-tailed models, the four approaches just mentioned all fall under the same category: heavy-tailed alternatives to the normal. The main difference is in the tail decay, from exponential to an absence of decay (see \autoref{fig:comparisons_PDFs}). The focus in this paper is on this classic category of approaches to robustness, and more precisely on the four aforementioned approaches. In \autoref{sec:overview}, we provide an overview of these, which will allow to make more precise what has been discussed above.

\normalsize

A frequentist thus has access to all these four approaches, while a traditional Bayesian only to three. Another contribution of this paper is to highlight that a way to close this gap and completely reconcile the Bayesian and frequentist robust approaches is to accept the use of improper models in Bayesian analyses \citep{jewson2022general}, while using in return proper prior distributions in order to obtain proper posterior distributions. This strategy in fact fits within the generalized Bayesian framework of \cite{bissiri2016general} under which it is acknowledged that models like normal linear regression are plausibly misspecified; the likelihood function is thus replaced by a loss function in a generalized posterior distribution. The complete reconciliation between Bayesian and frequentist approaches represents an argument in favour of adopting this framework, at least in the context of robustness against outliers. In \autoref{sec:Tukeymodel}, we provide an overview of the generalized Bayesian framework and consider Tukey's biweight improper model as an example of a case where it applies. Tukey's biweight M-estimator being a popular frequentist robust approach, it is important to study its generalized Bayesian counterpart. The final contribution of this paper is to provide an extensive theoretical study of the generalized Bayesian counterpart, with a robustness characterization result, a strong consistency result and a Bernstein--von Mises result, the latter allowing to calibrate the generalized posterior distribution for meaningful uncertainty quantification, which is often a challenge with such a distribution.

The notion of improper models has been introduced in \cite{jewson2022general}. By leveraging the generalized Bayesian framework of \cite{bissiri2016general}, a principled way to perform Bayesian inference for such models is proposed, including model selection. Tukey's biweight improper model is a motivating application, in particular, for comparison with the normal linear regression model to evaluate whether the improper heavy-tailed model is justified for the data set at hand. The method is based on the Hyvärinen score \citep{hyvarinen2005estimation}, whose evaluation for a model does not involve the normalizing constant of the distribution. This score can thus be evaluated for improper models for which the normalizing constant is not well defined. The proposed method is also relevant for variable selection and data-driven specification of tuning parameters. These aspects are not explored in this paper. We thus refer the reader to \cite{jewson2022general} for the details about their method and an interesting discussion about improper models. In light of the above, the focus of the work of \cite{jewson2022general} is different than ours and thus their work has virtually no overlap with ours; their study of Tukey's biweight improper model and ours are complementary.

\subsection{First example}\label{sec:example1}

\begin{figure}[ht]
 \centering\small
 $\begin{array}{cc}
   \includegraphics[width=0.5\textwidth]{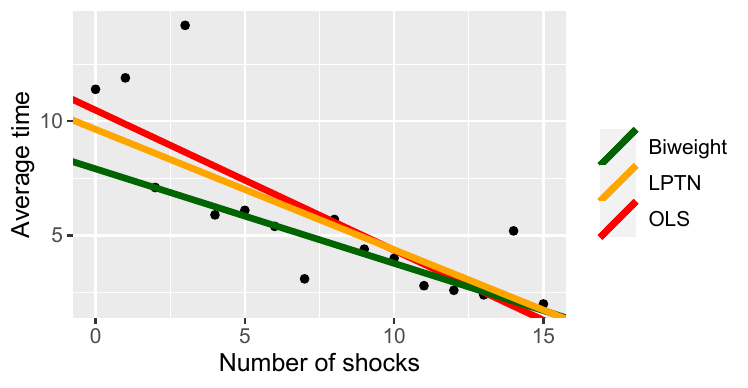} & \hspace{-5mm} \includegraphics[width=0.5\textwidth]{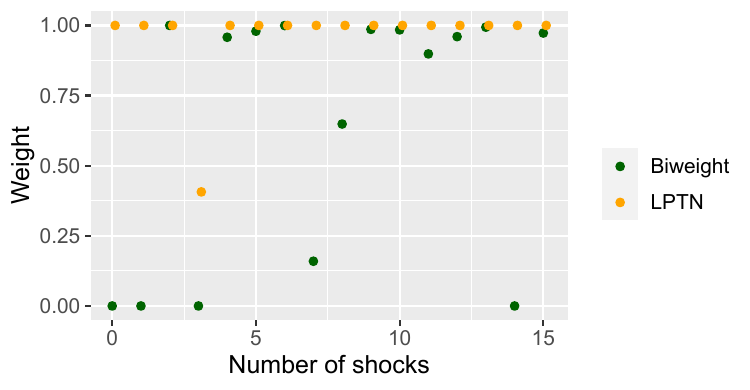} \cr
   \hspace{-10mm} \textbf{(a) } & \hspace{-14mm} \textbf{(b)}
  \end{array}$
  \vspace{-3mm}
\caption{(a) Estimation of a simple linear regression based on the \textsf{shock} data set using OLS and Tukey's biweight M-estimator, as well as the maximum a posteriori estimate of a Bayesian LPTN model. (b) Weight assigned to each data point in Tukey's biweight M-estimation and Bayesian LPTN model estimation.}\label{fig:scatter_ex_1}
\end{figure}

We finish this section with a first example where a significant difference is observed between estimation results obtained by a frequentist and a traditional Bayesian practitioner. The code to reproduce our numerical results is available online (see ancillary files on \url{https://arxiv.org/abs/2408.10478}). In \autoref{fig:scatter_ex_1} (a), we show estimation results for a simple linear regression, based on the \textsf{shock} data set accessible via the \textsf{R} package \textsf{RobStatTM} \citep{RobStatTM}. This data set is about an experiment conducted on rats to evaluate the average time to go through a shuttlebox depending on the number of electric shocks dispensed. This data set is presented and analysed in the book of \cite{maronna2019robust} (Chapter 4), a classic in the frequentist robustness literature.

We observe, as in \cite{maronna2019robust}, that Tukey's biweight M-estimation offers a better fit, in the sense that the regression line passes through the majority of the data points (which can be considered as the non-outliers); OLS estimation and Bayesian LPTN model estimation are quite influenced by the outliers. Huber M-estimation is in between OLS estimation and Bayesian LPTN model estimation. We evaluated a spectrum of tail decay for the LPTN model by trying different values for $\lambda$, and in particular, values close to 1, representing the properness limit of the LPTN model. The results are all similar to those in \autoref{fig:scatter_ex_1} (a). The tuning parameter in Tukey's biweight M-estimator is set to reach a 95-percent efficiency (the details follow in \autoref{sec:overview}).

The relationship between the dependent variable and the covariate in this example is somewhat complex. At first glance, one may see a decreasing trend which is non-linear at the beginning and then linear. However, it is not all data points with a small $x$ value that are in the upper left corner in \autoref{fig:scatter_ex_1} (a) and aligned with the non-linear trend (e.g., that with $x = 3$ is not). Also, based on the regression line estimated via Tukey's biweight M-estimator, there are data points with higher $x$ values that do not follow the linear trend (e.g., that with $x = 14$), at least not with the same error scale. An analysis such as that based on Tukey's biweight M-estimator in this example allows to gain perspective on the trends in the data set, which may motivate the practitioner to push further the analysis by using, for instance, a non-linear and heteroscedastic model.

In this example, the outlier contamination is severe, in the sense that there are many outliers, relatively to the number of data points and of unknown parameters, and the outliers are not so far away. In \autoref{sec:example_reserve}, we show that this example is not unique. We provide another real-data example where a significant difference is observed between estimation results obtained by a frequentist and a traditional Bayesian practitioner. This time the model is more complex as the number of covariates is larger, but the situation is similar in the sense that the outlier contamination is again severe. The simplicity of the example in this section allows to clearly visualize the trends in the data estimated by the different methods. With this simple example in mind, one can imagine that the difference in estimated trends is similar in higher dimensional problems like that in \autoref{sec:example_reserve}, but we cannot visualize it as clearly (explaining why we instead rely on objects like the residuals for visualization).

Significant differences between the most heavy-tailed frequentist and traditional Bayesian approaches are not expected in typical outlier contamination scenarios. In these scenarios, the difference between the Bayesian posterior distribution that excludes the outliers and the one that does not is not significant. There is no guarantee that it is the case in scenarios where the outlier contamination is severe, even though the LPTN model is wholly robust as this property is asymptotic. In such a scenario, Tukey's biweight M-estimator may be less influenced by the outliers due to the constant tails of the loss function and we may observe a significant difference between the most heavy-tailed frequentist and traditional Bayesian robust approaches. The influence of the outliers or any data point can be deduced from the assigned weight in the estimation; a formal definition of the weight function is given in \autoref{sec:info_M_Bayes}. In \autoref{fig:scatter_ex_1} (b), we present the weight assigned to each data point in Tukey's biweight M-estimation and Bayesian LPTN one. Note that a weight of 1 is assigned to all data points in OLS estimation.

Under the generalized Bayesian framework, Tukey's biweight improper model studied in \autoref{sec:Tukeymodel} leads to essentially the same estimated regression line as its M-estimator counterpart, with maximum a posteriori (MAP) estimates of $7.91$ and $ -0.41$ for the intercept and slope, respectively. The 95\% highest posterior density (HPD) credible intervals (CIs) are $(7.16, 8.65)$ and $(-0.49, -0.34)$. These MAP estimates and CIs are similar to those obtained with the normal model based on the data set excluding the (identified) outliers. Based on the whole data set, the estimates with the normal model are $10.48 \, (8.44, 12.54)$ and $ -0.61 \, (-0.85, -0.38)$. The MAP estimates lead to a regression line which is similar to that with OLS in \autoref{fig:scatter_ex_1} (a). Throughout, the prior distribution used for the regression coefficients is a weakly informative normal.

 In this example, Student's $t$ models with degrees of freedom around 1 manage to handle the outliers and allow to obtain similar estimation results to Tukey's biweight M-estimator. This is somewhat unexpected as the Student's $t$ PDF as a faster tail decay than the LPTN PDF. In situations where the outlier contamination is expected to be severe, practitioners may not even try Student's $t$ models as LPTN models are expected to be more robust. Also, to avoid low efficiency estimators, Student's $t$ models with such small degrees of freedom may not be considered \citep{gagnon2023theoretical}. It is typically recommended to use a Student's $t$ model with $\nu = 4$ degrees of freedom (displayed in \autoref{fig:comparisons_PDFs}) or similar, which has an efficiency around 90\%. In other words, practitioners may not observe such similar estimation results to Tukey's biweight M-estimator. In the example of \autoref{sec:example_reserve}, the expected scenario occurs and the estimation results with the LPTN model are less influenced by the presence of outliers than those with the Student's $t$ models (yet, the influence is significant, again contrarily as with Tukey's biweight M-estimator and its generalized Bayesian counterpart).

 Between the Student's $t$ and LPTN models, the difference in the tails is slight when the degrees of freedom of the former are small, but there is also a slight difference in PDF shape. The LPTN PDF exactly matching the standard normal on the central part, it decreases faster than the standard normal for a short interval at the beginning of the tails after which it goes above, a consequence of the continuity of the LPTN PDF with the constraint of integrating to 1. The Student's $t$ model is constructed otherwise with a PDF that becomes flatter when decreasing the degrees of freedom, which turns out to be an advantage in this example. The insensitivity of the estimation results with Student's $t$ models having degrees of freedom around 1 in this example made us realized that, if we could slightly flatten the LPTN PDF, we may end up with insensitivity in this example with this model as well. We managed to achieve this by going beyond the constraint $\lambda > 1$ and setting $\lambda = 1$ but changing nothing else in the model, thus yielding an improper distribution. This highlights that properness is a real limit for the Bayesian LPTN model in this example.

\normalsize

\section{Frequentist and traditional Bayesian approaches}\label{sec:overview}

 \subsection{M-estimators}\label{sec:M-estimators}

In linear regression, it is assumed that $y_1, \ldots, y_n$ are realizations of $n$ random variables $Y_1, \ldots, Y_n$ defined through the following model:
\begin{align}\label{eqn:linear_reg}
 Y_i = \bx_i^T \bbeta + \sigma \varepsilon_i, \quad i = 1, \ldots, n,
\end{align}
where $\bbeta =(\beta_1, \ldots, \beta_p)^T \in \re^p$ is a vector of regression coefficients, $\sigma > 0$ is a scale parameter and $\varepsilon_1, \ldots, \varepsilon_n \in \re$ are standardized errors. In an homoscedastic model, it is assumed that $\varepsilon_1, \ldots, \varepsilon_n$ are independent and identically distributed (IID) random variables, each having a PDF denoted here by $f$. Note that the covariates are not modelled and the vectors $\bx_1, \ldots, \bx_n$ are thus not considered as realizations of random variables, contrarily to $y_1, \ldots, y_n$.

The log-likelihood function is given by
\begin{align}\label{eqn:logL}
 \ell(\bbeta, \sigma) := -n \log\sigma + \sum_{i=1}^n \log f\left(\frac{y_i - \bx_i^T \bbeta}{\sigma}\right), \quad \bbeta \in \re^p, \sigma > 0.
\end{align}
When $f$ is the standard normal PDF, we have $f(\varepsilon) = g(\varepsilon) / m$ with $g(\varepsilon) = \exp(-\varepsilon^2/2)$ and $m = \sqrt{2 \pi}$, the normalizing constant. The quadratic term resulting from $-\log g$ produces extreme values when some residuals $y_i - \bx_i^T \bbeta$ are extreme, which is the case for outliers when the log-likelihood function is evaluated at parameter values reflecting the trend in the bulk of the data. To alleviate the impact and thus the sensitivity to outliers, the quadratic function is replaced by one that grows less rapidly in M-estimation. We can view this approach as replacing the function $g$. From the new function $g$, we can identify a log-likelihood function as in \eqref{eqn:logL} and consider M-estimation as maximum likelihood estimation of the model in \eqref{eqn:linear_reg}, but with this new function $g$. When this function is integrable, $g$ is equivalent to a PDF $f$, up to a normalizing constant $m$. When it is not integrable and $m$ is thus not well defined, we will still consider that \eqref{eqn:linear_reg} with this function $g$ is a model, but an improper one, as in \cite{jewson2022general}. Note that, from an M-estimation perspective, improperness is not a problem as the estimation procedure corresponds to the maximization of \eqref{eqn:logL} with $m$ omitted.

The Huber M-estimator results from setting
\begin{align*} % \label{eqn:huber}
-\log g(\varepsilon)=\left\{
\begin{array}{lcc}
                                                      \varepsilon^2 / 2  & \text{ if } & |\varepsilon|\leq k, \cr
                                                      k|\varepsilon| - k^2 / 2 & \text{ if } & |\varepsilon|>k, \cr
\end{array}
\right.
\end{align*}
with $k > 0$ a tuning parameter controlling the tradeoff between robustness and efficiency.\footnote{With the Huber M-estimator, the commonly used value of $k = 1.345$ allows the estimator to reach a 95-percent efficiency. The associated PDF is displayed in \autoref{fig:comparisons_PDFs}.} M-estimation in this case can thus be viewed as maximum likelihood estimation of the model in \eqref{eqn:linear_reg} with $f$ as follows:
 \begin{align}\label{eqn:f_Huber}
f(\varepsilon) = g(\varepsilon) / m =\left\{
\begin{array}{lcc}
                                                      \exp(-\varepsilon^2 / 2) / m  & \text{ if } & |\varepsilon|\leq k, \cr
                                                     \exp(-k|\varepsilon| - k^2 / 2) / m  & \text{ if } & |\varepsilon|>k, \cr
\end{array}
\right.
\end{align}
where $m = 2 \e^{-k^2 / 2} / k + \sqrt{2 \pi} (2 \Phi(k) - 1)$, $\Phi$ being the cumulative distribution function of the standard normal.

Given that this model is proper, the scale parameter $\sigma$ can be estimated through the maximization of \eqref{eqn:logL}. In the frequentist literature, it is often estimated otherwise \citep[Section 4.4]{maronna2019robust}. A strongly consistent robust estimator is typically used, where the estimator converges with probability 1 to the true scale parameter in the case where the true data-generating process is assumed to be the linear regression model (with regularity conditions on the error distribution). A normalized mean absolute deviation (MAD) of the residuals is often used, where the residuals are calculated using a $L_1$ estimator of $\bbeta$. The latter corresponds to the maximization of \eqref{eqn:logL} with respect to $\bbeta$ with $-\log g(\varepsilon) = |\varepsilon|$, which holds for any $\sigma >0$, like in the normal model with $-\log g(\varepsilon) = \varepsilon^2 / 2$ (and thus does not require estimating a scale). Using $-\log g(\varepsilon) = |\varepsilon|$ is equivalent to a Laplace model and is thus aligned with the tails of $f$ in \eqref{eqn:f_Huber}. The estimation is thus a two-step procedure where the first step is to estimate $\sigma$ as just described and the second one is to estimate $\bbeta$ through the maximization of \eqref{eqn:logL} with $f$ as in \eqref{eqn:f_Huber}. An advantage of Huber M-estimation is that the resulting function $\ell$ (recall \eqref{eqn:logL}) is concave and smooth, like with the normal errors, which is a desirable property for optimization purposes. A disadvantage is that the growth of $-\log g$ is still rapid, implying sensitivity to outliers, as observed in \autoref{sec:example1}.

Tukey's biweight M-estimator comes with a gain on one side, but a loss on the other. It results from setting
\begin{align}\label{eqn:biweight}
-\log g(\varepsilon)=\left\{
\begin{array}{lcc}
                                                      1-(1-(\varepsilon/k)^2)^3  & \text{ if } & |\varepsilon|\leq k, \cr
                                                      1 & \text{ if } & |\varepsilon|>k,
\end{array}
\right.
\end{align}
where we used the same notation $k$ as with the Huber M-estimator given that this tuning parameter plays the same role here. The numerical results of Sections \ref{sec:example1} and \ref{sec:example_reserve} have been obtained using the commonly used value of $k = 4.685$ which allows the estimator to reach a 95-percent efficiency (see Section 2.2 in \cite{maronna2019robust} for a definition of efficiency).  The function $-\log g$ in \eqref{eqn:biweight} has appealing properties. It is similar to the quadratic function on the central part (but it operates on a different scale), while being constant on the extremities and being smooth; see \autoref{sec:info_M_Bayes} for a figure. However, the resulting function $\ell$ is not concave.

The fact that $g$ is constant on the extremities implies that $g$ is not integrable and thus that $m$ is not well defined in this case. The model is thus improper and the scale parameter $\sigma$ cannot be estimated through the maximization of \eqref{eqn:logL}. Indeed, the function in \eqref{eqn:logL} diverges as $\sigma \rightarrow 0$ for any values of the residuals $y_i - \bx_i^T \bbeta$ as a result of the boundedness of $\log g$. The scale parameter can be estimated through the normalized MAD of the residuals, as mentioned above, with the residuals calculated using the $L_1$ estimator of $\bbeta$. Such an estimator of $\sigma$ is however sensitive to outliers due to the use of the $L_1$ estimator of $\bbeta$. A more robust alternative consists of solving simultaneously two equations, where the first one corresponds to a zero of the derivative of \eqref{eqn:logL} with respect to $\bbeta$ and the second one corresponds to the normalized MAD (written as an equation). The resulting Tukey's biweight M-estimator, that we denote by  $(\hat{\bbeta}_{\text{TM}}, \hat{\sigma}_{\text{TM}})$, is strongly consistent in the aforementioned sense \citep{maronna1981asymptotic}. It is used in the function \textsf{rlm} of the \textsf{R} package \textsf{MASS} \citep{MASS}.

%With the function $g$ in \eqref{eqn:biweight}, $g(\varepsilon) = \exp(-1)$ whenever $|\varepsilon|>k$, implying that this density assigns the same measure $\int_I g(\varepsilon) \, \d\varepsilon$ to all intervals $I$ of same length, regardless of the distance to the origin, as long as the condition $|\varepsilon|>k$ is satisfied for all points in the intervals. Viewing probability distributions as measures, this characteristic of Tukey's biweight M-estimator allows to gain understanding regarding its behaviour, even if it is associated to an improper model. All proper (monotonic) models assign decreasing probabilities $\int_I g(\varepsilon) \, \d\varepsilon$ with the distance to the origin. Regarding now the function to optimize, with a form as in \eqref{eqn:logL} but with $m$ omitted,
With Tukey's biweight loss function in \eqref{eqn:biweight}, we have that
\begin{equation}\label{eqn:tail_tukey}
 \log g\left(\frac{y_i - \bx_i^T \bbeta}{\sigma}\right) = -1 \quad \text{if} \quad |y_i - \bx_i^T \bbeta| / \sigma > k.
\end{equation}
Therefore, $\log g((y_i - \bx_i^T \bbeta) / \sigma)$ acts as a constant and does not actually vary as a function of $\bbeta$ and $\sigma$ in regions where $ |y_i - \bx_i^T \bbeta| / \sigma > k$; it is thus expected that outlier terms act as constants in regions reflecting the trend in the bulk of the data, implying that they are expected to be effectively ignored in these regions. The same observations can be made by analyzing the weight function (recall \autoref{fig:scatter_ex_1} (b) and that a formal definition of the weight function is given in \autoref{sec:info_M_Bayes}). In particular, Tukey's biweight M-estimator assign 0 weight to outliers with $ |y_i - \bx_i^T \bbeta| / \sigma > k$.

\begin{Remark}
 Tukey's biweight loss function is in fact defined in \cite{beaton1974fitting} through its derivative. Therefore, it is not unique. In \cite{maronna2019robust}, it is presented in the same form as in \eqref{eqn:biweight}. In other references, the form is different \citep{rousseeuw2003robust, jewson2022general}:
    \begin{align}\label{eqn:biweight2}
-\log g(\varepsilon)=\left\{
\begin{array}{lcc}
                                                      \frac{1}{2} \log(2 \pi) + \frac{\varepsilon^2}{2} - \frac{\varepsilon^4}{2 k^2} + \frac{\varepsilon^6}{6 k^4}  & \text{ if } & |\varepsilon|\leq k, \cr
                                                      \frac{1}{2} \log(2 \pi) + \frac{k^2}{6} & \text{ if } & |\varepsilon|>k.
\end{array}
\right.
\end{align}
The form adopted here is easier to connect to the weight function, whereas that in \eqref{eqn:biweight2} is easier to connect to the normal distribution.
\end{Remark}

\subsection{Heavy-tailed models in traditional Bayesian approaches}\label{sec:heavy-tailed}

In linear regression, the traditional Bayesian model is like in \eqref{eqn:linear_reg}, with the difference that the unknown parameters $\bbeta$ and $\sigma$ are considered random. We thus need to assume a dependence structure among all the random variables involved in the model, that is $\bbeta$, $\sigma$ and $\varepsilon_1, \ldots, \varepsilon_n$. Additionally to the IID assumption on $\varepsilon_1, \ldots, \varepsilon_n$, it is typically assumed that these latter random variables are independent of $(\bbeta, \sigma)$. Therefore, $Y_1, \ldots, Y_n$ are conditionally independent (given $\bbeta$ and $\sigma$), and the conditional PDF of $Y_i$ can be written in terms of that of $\varepsilon_i$, that is $f$. The conditional PDF of $\bY := (Y_1, \ldots, Y_n)^T$ corresponds to the likelihood function and the latter is multiplied by the prior density of the parameters in the posterior density. Let us denote the (possibly improper) prior density by $\pi$. The posterior density is thus as follows:
 \begin{align}\label{eqn:posterior}
  \pi(\bbeta, \sigma \mid \by) := \frac{\pi(\bbeta, \sigma) \prod_{i=1}^n \frac{1}{\sigma} f\left(\frac{y_i - \bx_i^T \bbeta}{\sigma}\right)}{\int_0^\infty \int_{\re^p} \pi(\bbeta, \sigma) \prod_{i=1}^n \frac{1}{\sigma} f\left(\frac{y_i - \bx_i^T \bbeta}{\sigma}\right) \, \d\bbeta \, \d\sigma}, \quad \bbeta \in \re^p, \sigma > 0,
 \end{align}
 assuming that the integral in the denominator is finite. Note that the posterior distribution is conditional on $\bY = \by$ only as the vectors $\bx_1, \ldots, \bx_n$ are not considered as realizations of random variables.

In a proper probabilistic model, $\pi$ is a PDF. It is common that Bayesians accept the use of improper prior densities, which are typically used to reduce the impact of the prior on the posterior. Jeffreys priors, for instance, have this objective. In linear regression, the Jeffreys prior corresponds to $\pi(\bbeta, \sigma) \propto 1 / \sigma$. The numerical results of Sections \ref{sec:example1} and \ref{sec:example_reserve} for the Student's $t$ and LPTN models have been obtained using $\pi(\bbeta, \sigma) \propto 1$ so that the MAP estimate corresponds to the maximizer of \eqref{eqn:logL}, essentially as with the M-estimators in \autoref{sec:M-estimators}, the crucial difference being in the function $g$ used. Maximizing \eqref{eqn:logL} corresponds to maximum likelihood estimation which can also be viewed as M-estimation of the normal model when $-\log g(\varepsilon) \neq \varepsilon^2 / 2$. From this point of view, the weight functions can be derived for the Bayesian models as well. Note that when using a different prior density than $\pi(\bbeta, \sigma) \propto 1$, MAP estimation corresponds to \emph{regularized} maximum likelihood estimation \citep[Chapter 6]{james2013introduction}.

In Bayesian normal linear regression, $f$ is the standard normal PDF. As mentioned in \autoref{sec:connection}, the robustness problems of this model are caused by the exponential decay of the PDF which makes extremely unlikely the presence of outliers. When the likelihood function is evaluated at parameter values reflecting the trend in the bulk of the data, the light tails penalize heavily those values for the outliers, diminishing significantly the likelihood-function value. The analogous phenomenon arises when the likelihood function is evaluated at parameter values reflecting the trend in the outliers: those values are heavily penalized for the bulk of the data. As a result, parameter values between those aforementioned become more plausible, representing an undesirable compromise.

The idea of using heavy-tailed distributions comes from these considerations. A natural candidate is the Student's $t$ distribution given its resemblance to the normal. However, as mentioned in \autoref{sec:connection}, it is not able to provide whole robustness. The LPTN distribution does and also has a PDF similar to the normal. We now present this PDF and explain how it is able to reach whole robustness. While being a desirable property, whole robustness is, however, not a guarantee of insensitivity to outliers, as seen in \autoref{sec:example1} and as will be seen in \autoref{sec:example_reserve}. In \autoref{sec:Tukeymodel}, we explain how improper models such as Tukey's biweight can be used in (generalized) Bayesian analyses as potential approaches to gain robustness.

The LPTN PDF has an hyperparameter $\rho \in (2\Phi(1) - 1, 1) \approx (0.6827, 1)$ and is defined piecewise, like the densities associated with the Huber and Tukey's biweight M-estimators (recall \eqref{eqn:f_Huber} and \eqref{eqn:biweight}). More precisely, it is defined as
\begin{align}\label{eqn:LPTN}
f(\varepsilon)=\left\{
\begin{array}{lcc}
                                                      \varphi(\varepsilon)  & \text{ if } & |\varepsilon|\leq \tau, \cr
                                                      \varphi(\tau)\,\frac{\tau}{|\varepsilon|}\left(\frac{\log \tau}{\log |\varepsilon|}\right)^{\lambda} & \text{ if } & |\varepsilon|>\tau, \cr
\end{array}
\right.
\end{align}
where $\tau > 1$ and $\lambda > 1$ are functions of $\rho$ with
\begin{align*}
 & \tau=\Phi^{-1}((1+\rho)/2) = \{\tau : \Prob(-\tau \leq Z \leq \tau)= \rho \,\text{ for }\, Z\, \sim \, \mathcal{N}(0,1)\}, \cr
 & \lambda=1 + 2(1-\rho)^{-1}\varphi(\tau) \, \tau \log(\tau),
 \end{align*}
$\varphi$ being the standard normal PDF. The hyperparameter $\rho$ controls the tradeoff between robustness and efficiency. The results shown in \autoref{fig:scatter_ex_1} were obtained with $\rho = 0.9$ (PDF displayed in \autoref{fig:comparisons_PDFs}), which allows to reach a good tradeoff \citep{gagnon2020}. By choosing $\rho$ close to the lower bound of the admissible values, that is $2\Phi(1) - 1$, we obtain a PDF with $\lambda$ close to 1, which is the limit in terms of tail heaviness of this model. The results regarding the \textsf{shock} data set with such small $\rho$ are not significantly different from those with $\rho = 0.9$. Note that, if we consider $\rho$ as a parameter of the model, like $\bbeta$ and $\sigma$, a value close to $0.9$ maximizes the posterior density in this example.

In \textit{resolution of conflict},  the outliers and the non-outliers are seen as two sources which suggest plausible regions for the parameters that are significantly different \citep{o2012bayesian}. This line of research aims to establish if a model allows for an effective resolution of conflict and, if so, to provide a mathematical characterization of it. For several decades, authors have worked on characterizing the robustness of simple models, such as the location model (e.g., \cite{dawid1973posterior}, \cite{o1979outlier} and \cite{angers2007conflicting}) and the location--scale model (e.g., \cite{andrade2011bayesian} and \cite{desgagne2015robustness}). The last decade have witnessed significant progress in this line of research, with a characterization in linear regression \citep{DesGag2019, gagnon2020, gagnon2020PCR, 10.1214/22-BA1330, gagnon2023theoretical, hamura2020log, HAMURA2024110130, hamura2024short}, generalized linear models \citep{gagnon2023GLM, hamura2021robust}, Student's $t$ process \citep{https://doi.org/10.1111/sjos.12611}, accelerated failure time models \citep{hamura2025robust} and multivariate linear models \citep{hamura2025multivariate}.

In this line of research, the resolution of conflict and robustness are characterized through the asymptotic behaviour of the posterior distribution, under a framework where the outliers move away from the bulk of the data. Under such a framework, the (observed and non-random) outliers $(\bx_i, y_i)$ are indexed by a variable $\vartheta \in \na$ and move away from the bulk of the data as $\vartheta \rightarrow \infty$ in a mathematically precise sense. We thus obtain a sequence of (observed and non-random) data sets and posterior distributions indexed by $\vartheta$. For each fixed $\vartheta$, the posterior distribution is defined as in \eqref{eqn:posterior}. An observed data set with outliers can be thought of as an observed data set with a large $\vartheta$ and the theoretical results obtained under the asymptotic framework allow to characterize the posterior distribution under such a large $\vartheta$. In linear regression, in the articles referenced above, it is considered that the outliers move along particular paths for technical reasons. It is considered that the vectors $\bx_i$ are fixed as $\vartheta \rightarrow \infty$, that is they are independent of $\vartheta$ (which does not prevent them from being extreme), and the observations $y_i \rightarrow \pm\infty$. In \autoref{sec:robustness}, we leverage the boundedness of Tukey's biweight loss function to obtain a theoretical result under a more general asymptotic framework.

From \eqref{eqn:posterior}, we understand that the limiting behaviour of the conditional PDF of $Y_i$ evaluated at an outlying point is central to the characterization of the robustness. When $f$ is an LPTN, we have that, for any fixed $\bbeta \in \re^p$, $\sigma > 0 $ and $\rho \in (2\Phi(1) - 1, 1)$ (which implies that $\tau$ and $\lambda$ are also fixed),
 \begin{align}\label{eq:limit_LPTN}
  \lim_{y_i \rightarrow \pm \infty} \frac{\frac{1}{\sigma} f\left(\frac{y_i - \bx_i^T \bbeta}{\sigma}\right)}{f(y_i)} = \lim_{y_i \rightarrow \pm \infty} \frac{|y_i|}{|y_i - \bx_i^T \bbeta|} \left(\frac{\log|y_i|}{\log(|y_i - \bx_i^T \bbeta| / \sigma)}\right)^\lambda = 1.
 \end{align}
 This result suggests that the PDF term of an outlier in the likelihood function or the posterior density behaves in the limit like $f(y_i)$. When considering $\rho$ as an hyperparameter and thus fixed, $f(y_i)$ acts as a constant in these functions (of $\bbeta$ and $\sigma$). It is thus expected that outliers are wholly excluded of these functions when normalized; this is proved in \cite{gagnon2020}, guaranteeing that the LPTN is \textit{wholly robust}. We observe that the reason why the limit in \eqref{eq:limit_LPTN} holds for the PDF in \eqref{eqn:LPTN} with tails $f(\varepsilon) \propto (1 /|\varepsilon|) (1 / (\log |\varepsilon|)^{\lambda})$ is the power of $1$ in the term $1/|\varepsilon|$, combined with the log term. We thus understand that if the power was larger than $1$, as with the Student's $t$ PDF, the limit would not be 1. It is indeed given by $\sigma^\nu$, where $\nu > 0$ represents the degrees of freedom. For this model, the terms $(1 / \sigma) f((y_i - \bx_i^T \bbeta) / \sigma)$ in the posterior density thus asymptotically behaves like $f(y_i) \, \sigma^\nu\propto \sigma^\nu$. Therefore, in the limiting posterior density, we do get rid of terms depending on $y_i$ and thus of $\vartheta$, which can be seen as the source of outlyingness, but there is a trace of the outliers in the posterior distribution, namely $\sigma^\nu$ for each outlier. The Student's $t$ model is thus \textit{partially robust}.

 The limit to a constant with the LPTN PDF in \eqref{eq:limit_LPTN} contrasts with the equality to a constant beyond a threshold with Tukey's biweight loss function in \eqref{eqn:tail_tukey}.  The speed at which the limiting regime with the LPTN PDF is reached dictates how far the outliers need to be in order to be effectively ignored. With Tukey's biweight loss function, the outliers only need to be such that $ |y_i - \bx_i^T \bbeta| / \sigma > k$ to be effectively ignored. Contrarily to Tukey's biweight M-estimator, the LPTN model can indeed only assign a weight to outliers that tends to 0. These observations allow for a clear explanation of the difference between the estimation results shown in Sections \ref{sec:example1} and \ref{sec:example_reserve}. We present a proposition stating that the tail behaviour of Tukey's biweight loss function in \eqref{eqn:tail_tukey} is unattainable for a proper model and thus a limiting result as in \eqref{eq:limit_LPTN} is the best one can hope for.

 \begin{Proposition}\label{prop:impossible}
  Let $f: \re \rightarrow (0, \infty)$ be a strictly positive PDF. Then, $f$ cannot be constant beyond a threshold $k > 0$, regardless of the value of $k$. 
\end{Proposition}

\begin{proof} Trivial. \end{proof}

 Using heavy-tailed alternatives to the normal model comes with a gain in robustness, but it also comes with an increased computational complexity. For computing estimates, Markov chain Monte Carlo (MCMC) represents the go-to approach. Parameter estimates can be computed using, for instance, Hamiltonian Monte Carlo \citep{Duane1987, neal2011mcmc}. Variable selection and model averaging can be performed using reversible jump algorithms \citep{green1995reversible, green2003trans}; see \cite{gagnon2019RJ}, \cite{gagnon2019NRJ} and \cite{gagnon2023} for efficient informed and non-reversible variants.

 \section{Generalized Bayes: Case of Tukey's biweight model}\label{sec:Tukeymodel}

 In this section, we start with  an overview of the generalized Bayesian framework of \cite{bissiri2016general} in \autoref{sec:generalizedBayes}. As mentioned previously, an important characteristic of this framework is that it allows for improper models. We finish the sub-section by presenting the generalized posterior distribution associated with Tukey's biweight improper model. We next turn to a theoretical study of the latter, with a characterization of the robustness in \autoref{sec:robustness} and a large-sample asymptotic result in \autoref{sec:largesample}. The proofs of the theoretical results are deferred to \autoref{sec:proofs}. To our knowledge, it is the first time that results describing the properties of a robust improper model appear. We note that large-sample asymptotic results for classical robust Bayesian heavy-tailed models are rare. The only result we are aware of is Theorem 2 in \cite{gagnon2023theoretical} about the Student's $t$ model. Given the connection between M-estimators and heavy-tailed models, the recent and interesting work of \cite{marusic2025theoretical} can be seen as providing large-sample asymptotic results for robust Bayesian heavy-tailed models. This work introduces generic large-sample asymptotic results and robustness results for what is referred to as the \textit{M-posterior}, being a generic generalized posterior distribution derived from a generic M-estimator.

 \subsection{Generalized Bayesian framework}\label{sec:generalizedBayes}

  In a generalized Bayesian framework, \cite{bissiri2016general} investigate the following question: in a situation where a prior belief about a parameter of interest $\btheta \in \re^d$ is expressed through a distribution $\pi$ and this belief is connected to observations $(y_1, \ldots, y_n)^T = \by$ via a loss function $l$ (not necessarily derived from a probabilistic model), is there a coherent procedure to update the prior belief, thus leading to a posterior? The authors provide an affirmative answer by showing that a coherent procedure (in a sense made precise in their paper) is to calculate the generalized posterior density as follows:
 \[
  \pi(\btheta \mid \by) \propto \exp(-l(\btheta, \by)) \, \pi(\btheta), \quad \btheta \in \re^d.
 \]
   In the case where $l = - \ell$ with $\ell$ as in \eqref{eqn:logL} and derived from a proper probabilistic model, the generalized posterior density corresponds to the usual posterior density as in \eqref{eqn:posterior}, showing that this framework is indeed a generalized Bayesian one. In this framework, we are allowed to set $l = - \ell$ as in \eqref{eqn:logL} with Tukey's biweight loss function corresponding to $-\log g$ in \eqref{eqn:biweight} (but with $m$ omitted), as highlighted in \cite{jewson2022general}. Using an improper density $g$, but in combination with a proper prior distribution in order to obtain a proper (generalized) posterior distribution, may seem reasonable for a Bayesian as it is symmetrical to using an improper prior density in combination with a proper model. However, the requirement of using a proper prior distribution comes with the risk of misspecifying the prior, meaning specifying a prior which then enters in conflict with the trend in the data. This may result in skewed inference and predictions.

 By experimenting with Tukey's biweight improper model, two challenges arose. Firstly, including the scale parameter $\sigma$ as a parameter of interest in the generalized posterior distribution led to instability for the reason mentioned in \autoref{sec:M-estimators}. To address this, we propose a pragmatic two-step approach where $\sigma$ is first estimated by $\hat{\sigma}$ and then a generalized posterior distribution is derived for $\bbeta$. We propose to use Tukey's biweight M-estimator, that is $\hat{\sigma} = \hat{\sigma}_{\text{TM}}$, corresponding to the solution of the two equations mentioned in \autoref{sec:M-estimators} (resulting in $(\hat{\bbeta}_{\text{TM}}, \hat{\sigma}_{\text{TM}})$). % Below, it will be made clear that the generalized MAP estimate of $\bbeta$ is expected to be similar to $\hat{\bbeta}_{\text{TM}}$ when the prior distribution is not in conflict with $\hat{\bbeta}_{\text{TM}}$ or when the prior distribution is weakly informative.

 The second challenge was that uncertainty quantification was not meaningful if we directly set $l = - \ell$ with $\ell$ as in \eqref{eqn:logL} and Tukey's biweight loss function corresponding to $-\log g$ in \eqref{eqn:biweight} (but with $m$ omitted); the CIs were for instance overly large. Both challenges are due to $g$ being not a proper density function. A solution often employed within the generalized Bayesian framework to obtain meaningful uncertainty quantification is to include a \textit{tempering parameter} $w > 0$ in the generalized posterior distribution for calibration \citep{holmes2017assigning}. For Tukey's biweight improper model, we thus propose to use the following generalized posterior density:
 \begin{align}\label{eq:postTukey}
  \pi(\bbeta \mid \by) &\propto \exp\left(-w \left(n \log\hat{\sigma}_{\text{TM}} - \sum_{i=1}^n \log g\left(\frac{y_i - \bx_i^T \bbeta}{\hat{\sigma}_{\text{TM}}}\right)\right)\right) \, \pi(\bbeta) \\
   &= \pi(\bbeta) \left[\prod_{i=1}^n \frac{1}{\hat{\sigma}_{\text{TM}}} g\left(\frac{y_i - \bx_i^T \bbeta}{\hat{\sigma}_{\text{TM}}}\right)\right]^w, \quad \bbeta \in \re^p, \nonumber
 \end{align}
 with $-\log g$ as in \eqref{eqn:biweight}. In \autoref{sec:largesample}, we present a Bernstein--von Mises result which indicates an asymptotically valid value for $w$ leading to meaningful uncertainty quantification. We propose to set $w$ to this value; we proceeded accordingly for the numerical experiments in Sections \ref{sec:example1} and \ref{sec:example_reserve}. Note that $\hat{\bbeta}_{\text{TM}}$ is a zero of the derivative of $\ell$ in \eqref{eqn:logL}, and thus of the function in the exponential in \eqref{eq:postTukey}. Therefore, the generalized MAP estimate of $\bbeta$ is expected to be similar to $\hat{\bbeta}_{\text{TM}}$, except when the prior information is in conflict with $\hat{\bbeta}_{\text{TM}}$ and suggests significantly different plausible values for $\bbeta$.

 The only remaining tuning parameter required to use Tukey's biweight improper model is $k$. We empirically observed that $k = 4.685$, allowing the M-estimator to reach a 95\% efficiency, leads to good results, both in M-estimation and generalized Bayesian estimation. We thus propose to set $k = 4.685$; this value was used for the numerical experiments in Sections \ref{sec:example1} and \ref{sec:example_reserve}. For the generalized Bayesian estimation, we use Hamiltonian Monte Carlo (HMC) to compute posterior means and CIs. In \autoref{sec:info_estimation}, we present the gradient of the log generalized posterior density, which is required for implementing HMC. Note that the method proposed in \cite{jewson2022general} allows the tuning parameters to be specified otherwise, namely in a data-driven manner, and for statistical inference on the tuning parameters.

 We finish this section with a proposition stating that the generalized posterior distribution associated with Tukey's biweight improper model is in general well defined.
 \begin{Proposition}\label{prop:proper}
   Suppose that the prior distribution $\pi$ is proper and that $\hat{\sigma}_{\text{TM}} \in (0, \infty)$ for the data set $\{\bx_i, y_i\}_{i=1}^n$ at hand. Therefore, the generalized posterior distribution defined through \eqref{eq:postTukey} is proper. Additionally, the moments of order $\kappa \in \re$ exist, provided that these moments exist under the prior distribution. 
\end{Proposition}

\subsection{Characterization of the robustness }\label{sec:robustness}

 In this section, we characterize the robustness of Tukey's biweight improper model under an asymptotic framework where outliers move away from the bulk of the data. As explained in \autoref{sec:heavy-tailed}, the outliers $(\bx_i(\vartheta), y_i(\vartheta))$ are indexed by a variable $\vartheta \in \na$ and move away from the bulk of the data as $\vartheta \rightarrow \infty$. The generalized posterior distribution $\pi_\vartheta(\, \cdot \mid \by)$ defined as in \eqref{eq:postTukey} is thus indexed in this sub-section by $\vartheta$ and we are interested in characterizing the behaviour of this distribution when $\vartheta$ is large. We present assumptions and notation, and next a result characterizing the robustness.

 To focus on robustness against outliers, we first assume that the prior $\pi$ is independent of $\vartheta$ (which explains why it is not indexed by $\vartheta$). This assumption reflects the situation where the prior distribution is not in conflict with the trend in the bulk of the data; see \cite{10.1214/22-BA1330} for a study of robustness of heavy-tailed prior distributions against conflicting prior information in regression. We also assume that the tuning parameter $k > 0$ in Tukey's biweight loss (recall \eqref{eqn:biweight}) is set by the user independently of $\vartheta$ as proposed in \autoref{sec:generalizedBayes}. We assume the same for the tempering parameter $w > 0$, meaning that it is set by the user independently of $\vartheta$, which is what we propose in \autoref{sec:largesample} and which is observed to be effective in the examples of Sections \ref{sec:example1} and \ref{sec:example_reserve}.

 \begin{Assumption}\label{ass:prior}
  The prior distribution $\pi$ is fixed and does not depend on $\vartheta$. Also, it is proper. The tuning parameters $k > 0$ and $w > 0$ are fixed and do not depend on $\vartheta$.
 \end{Assumption}

 We now present notation that allows us to state the second assumption. Let us denote the index set of the outliers $(\bx_i(\vartheta), y_i(\vartheta))$ by $\Oset \subset \{1, \ldots, n\}$. The number of outliers is thus given by the cardinality of $\Oset$, denoted by $|\Oset|$, and the index set of the non-outliers $(\bx_i, y_i)$ (that are fixed and independent of $\vartheta$) by $\Oset^\mathsf{c} := \{1, \ldots, n\} \setminus \Oset$.

 We use an assumption to represent the asymptotic framework where the outliers $(\bx_i(\vartheta), y_i(\vartheta))$ move away from the bulk of the data as $\vartheta \rightarrow \infty$. It essentially states that the absolute value of the residual $|y_i(\vartheta) -  \bx_i(\vartheta)^T \bbeta| \rightarrow \infty$ for all $\bbeta \in \re^p$, except on a set of Lebesgue measure 0. The asymptotic framework described in \autoref{sec:heavy-tailed}, which we refer to as \textit{outcome shift}, with $y_i(\vartheta) \rightarrow \pm \infty$ and $\bx_i(\vartheta)$ fixed as $\vartheta \rightarrow \infty$, is thus a special case. Another special case, which we refer to as \textit{covariate shift}, is that where one component of $\bx_i(\vartheta)$, say $x_{ij}(\vartheta)$, goes to $\pm\infty$ and all the other components of $\bx_i(\vartheta)$ and $y_i(\vartheta)$ are fixed and do not depend on $\vartheta$. In this case, $|y_i(\vartheta) -  \bx_i(\vartheta)^T \bbeta| \rightarrow \infty$ for all $\bbeta \in \re^p$ except when $\beta_j = 0$, which can be ignored as this has Lebesgue measure 0. The outliers are thus allowed to move away from the bulk of the data in different ways.

 \begin{Assumption}\label{ass:outliers}
  For all $i \in \Oset$ and $\bbeta \in A \subset \re^p$ with $A^\mathsf{c}$ of Lebesgue measure 0,
  \[
   |y_i(\vartheta) -  \bx_i(\vartheta)^T \bbeta| \rightarrow \infty \quad \text{as} \quad \vartheta \rightarrow \infty.
  \]
  For all $i \in \Oset^\mathsf{c}$, $(\bx_i, y_i)$ is fixed and does not depend on $\vartheta$.
 \end{Assumption}

 Under \autoref{ass:outliers}, Tukey's biweight loss function $-\log g$ in \eqref{eqn:biweight} is such that, for large enough $\vartheta$,
  \begin{align}\label{eqn:limitTukey}
 -\log  g\left(\frac{y_i(\vartheta) - \bx_i(\vartheta)^T \bbeta}{\hat{\sigma}}\right) = 1,
 \end{align}
 for all $i \in \Oset$, $\bbeta \in A$ and fixed $\hat{\sigma} \in (0, \infty)$, the latter representing in this sub-section a generic robust estimator of $\sigma$. Note that the reason we need to consider an asymptotic framework is specifically because we need \eqref{eqn:limitTukey} to hold for all $\bbeta \in A$. As a consequence of \eqref{eqn:limitTukey}, we expect the density terms of outliers in the generalized posterior density $\pi_\vartheta(\, \cdot \mid \by)$ to be excluded in the limiting generalized posterior density. The theoretical result stated below is about this. Therefore, with the generalized posterior density presented in \eqref{eq:postTukey} associated with Tukey's biweight improper model, whole robustness is expected, provided that $\hat\sigma$ is wholly robust. Whole robustness is not an aspect that has been studied in the frequentist literature. Thus, it has not been established whether the estimator $\hat{\sigma}_{\text{TM}}$ that we propose to use in \autoref{sec:generalizedBayes} is wholly robust. In terms of influence of outliers, the notion of \textit{breakdown point} is instead studied in the frequentist literature. It corresponds to the proportion of outliers that an estimator can handle, in the sense that the difference between the estimation with and without the outliers is bounded for any location of the outlying data points $(\bx_i(\vartheta), y_i(\vartheta))$ (see, e.g., Chapter 1 in \cite{rousseeuw2003robust}). It is beyond the scope of this paper to establish whether estimators like $\hat{\sigma}_{\text{TM}}$ are wholly robust.  In \autoref{sec:asymptotic_scale}, we present the results of a simple numerical experiment suggesting that, for $\hat{\sigma}_{\text{TM}}$, outliers have an impact asymptotically as $\vartheta \rightarrow \infty$, but it is not significant in some situations (e.g., when the proportion of outliers is not close to the breakdown point, see Chapter 3 in \cite{rousseeuw2003robust} for the breakdown point which depends on $k$). This in turn suggests that the estimator is close to being wholly robust in these situations.

 For the characterization of the robustness established in this sub-section, we thus assume that the number of outliers is such that the breakdown point of the robust estimator of $\sigma$ is not reached. We highlight a dependence of the estimator on $\vartheta$ by denoting it for the rest of the sub-section by $\hat\sigma_\vartheta$. In the situation where the breakdown point is not reached, we have that $\hat\sigma_\vartheta$ is finite for all $\vartheta$ and $\hat\sigma_\vartheta \rightarrow \bar\sigma$ as $\vartheta \rightarrow \infty$, where $\bar\sigma \in (0, \infty)$ is the limiting estimate. In the case where the estimator is wholly robust, $\bar\sigma$ corresponds to the estimate without the outliers.
 \begin{Assumption}\label{ass:estimator}
  We have that $\hat\sigma_\vartheta \in (0, \infty)$ for all $\vartheta$ and $\hat\sigma_\vartheta \rightarrow \bar\sigma \in (0, \infty)$ as $\vartheta \rightarrow \infty$.
 \end{Assumption}

 We now define $\pi(\, \cdot \mid \by_{\Oset^\mathsf{c}})$, where $\by_{\Oset^\mathsf{c}} = \{y_i: i \in \Oset^\mathsf{c}\}$, which will be seen to be the limiting generalized posterior density, highlighting that the outliers are excluded in the limit (at least when the robust estimator of $\sigma$ is wholly robust):
 \[
 \pi(\bbeta \mid \mathbf{y}_{\Oset^\mathsf{c}}) = \pi(\boldsymbol\beta) \left[\prod_{i \in \Oset^\mathsf{c}} \frac{1}{\bar{\sigma}} g\left(\frac{y_i - \bx_i^T \bbeta}{\bar{\sigma}}\right)\right]^w \Bigg/ m(\mathbf{y}_{\Oset^\mathsf{c}}), \quad \boldsymbol\beta \in \re^p,
\]
where
\begin{align*}%\label{eqn:constant_wo}
 m(\mathbf{y}_{\Oset^\mathsf{c}}) =  \int_{\re^p} \pi(\boldsymbol\beta) \left[\prod_{i \in \Oset^\mathsf{c}} \frac{1}{\bar\sigma} g\left(\frac{y_i - \bx_i^T \bbeta}{\bar\sigma}\right)\right]^w \d\boldsymbol\beta.
\end{align*}
Note that \autoref{prop:proper} also guarantees that $m(\mathbf{y}_{\Oset^\mathsf{c}}) < \infty$. Note also that $m_\vartheta(\by) < \infty$ for all $\vartheta$ under Assumptions \ref{ass:prior}-\ref{ass:estimator} by \autoref{prop:proper}, that is the normalizing constant of the generalized posterior density is finite, where
\[
 m_\vartheta(\by) =  \int_{\re^p} \pi(\boldsymbol\beta) \left[\prod_{i \in \Oset^\mathsf{c}} \frac{1}{\hat\sigma_\vartheta} g\left(\frac{y_i - \bx_i^T \bbeta}{\hat\sigma_\vartheta}\right)\right]^w \left[\prod_{i \in \Oset} \frac{1}{\hat\sigma_\vartheta} g\left(\frac{y_i(\vartheta) - \bx_i(\vartheta)^T \bbeta}{\hat\sigma_\vartheta}\right)\right]^w \d\boldsymbol\beta.
\]

We are now ready to present the main result of this sub-section, in which $\vartheta \rightarrow \infty$, implying that $|y_i(\vartheta) -  \bx_i(\vartheta)^T \bbeta| \rightarrow \infty$ for all $i \in \Oset$ and $\bbeta \in A \subset \re^p$, and $\hat\sigma_\vartheta \rightarrow \bar\sigma$, all the other components in $\pi_\vartheta(\, \cdot \mid \by)$ being fixed and independent of $\vartheta$.

\begin{Theorem}\label{Thm:robustness}
Suppose Assumptions \ref{ass:prior}-\ref{ass:estimator} hold. As $\vartheta \rightarrow \infty$,
 \begin{description}
   \item[(a)]
   \[
    \frac{m_\vartheta(\by)}{\hat\sigma_\vartheta^{-w|\Oset|} \exp(-w|\Oset|)} \rightarrow m(\by_{\Oset^\mathsf{c}});
\]
     \item[(b)] $\pi_\vartheta(\bbeta \mid \by) \rightarrow \pi(\bbeta \mid \by_{\Oset^\mathsf{c}})$, for all $\bbeta \in A$

\item[(c)] $\pi_\vartheta(\, \cdot \mid \by) \rightarrow \pi(\, \cdot \mid \by_{\Oset^\mathsf{c}})$ (convergence in distribution).
\end{description}
\end{Theorem}

An appealing aspect of \autoref{Thm:robustness} is that the assumptions are not restrictive and simple. The results are also simple. Result (a) states that $m_\vartheta(\by)$ is asymptotically equivalent to $m(\by_{\Oset^\mathsf{c}}) \, \hat\sigma_\vartheta^{-w|\Oset|} \exp(-w|\Oset|)$ (the last term coming from the equality in \eqref{eqn:limitTukey}). The proof essentially consists of showing that we are allowed to interchange the limit and the integral. As mentioned in \cite{hamura2024short} and \cite{gagnon2025simple}, such proofs are generally technical and not intuitive. It is however not the case with Tukey's biweight improper model, a consequence of the fact that $g$ is lower bounded.

Results (b) and (c) are obtained relatively easily from Result (a). Result (b) ensures, under some regularity conditions, the convergence of the MAP estimate and thus that the latter is robust. Result (c) indicates that any estimation of $\bbeta$ based on posterior quantiles (e.g., using posterior medians or CIs) is robust to outliers. In the proof of \autoref{Thm:robustness}, it is noted that the moments also converge and are thus robust, as long as these moments exist under the prior distribution. All these results characterize the limiting behaviour of a variety of Bayes estimators.

Note that, in the statement of \autoref{Thm:robustness}, there is no condition on the maximum number of outliers other than implicitly in \autoref{ass:estimator} (related to the breakdown point of the robust estimator of $\sigma$). This implies that it is solely the breakdown point of this estimator which dictates the breakdown point of Tukey's biweight improper model. Note that, in the situation where $\sigma$ is assumed known and fixed by the user (a situation that is not considered in this article) and all observations are outliers, the generalized posterior distribution converges to the prior distribution. In other words, the whole data set is asymptotically ignored in this situation, which is undesirable. This notion is discussed in \cite{marusic2025theoretical} within their framework of \textit{posterior breakdown point} in the greater generality of bounded loss functions and proper prior distributions.

 \subsection{Large-sample theory}\label{sec:largesample}

 In this section, we characterize the behaviour of the generalized posterior distribution under Tukey's biweight improper model when the sample size is large. We achieve this by establishing a theoretical result under an asymptotic framework where the data comes from a generic linear regression model and $n \rightarrow \infty$.

Contrarily to before, we assume that each $\bx_i := (x_{i1}, \ldots, x_{ip})^T$ is a realization of a random vector $\bX_i := (X_{i1}, \ldots, X_{ip})^T$. Thus, we assume that the true generating model is as follows: $\bZ_1 := (\bX_1, Y_1), \ldots, \bZ_n := (\bX_n, Y_n)$ are IID random variables such that $\bX_i \sim \mu_{\bX}$ and $Y_i = \bX_i^T \bbeta_0 + \sigma_0 \varepsilon_i$ with $\bbeta_0 := (\beta_{01}, \ldots, \beta_{0p})^T \in \re^p$ being the true (fixed) coefficient vector, $\sigma_0 > 0$ being the true (fixed) scale parameter, $\varepsilon_1, \ldots, \varepsilon_n$ being IID random variables (independent of $(\bX_1, \ldots, \bX_n)$) each having a PDF $f_0$  and $\mu_{\bX}$ being a probability distribution. In fact, we consider that all the components in $\bX_i$ are random, except the first one, $X_{i1}$, which is equal to 1 for the intercept. We thus consider that $\mu_{\bX}$ is the distribution of all the components in $\bX_i$, except the first one. We thus make an abuse of notation by writing $\bX_i \sim \mu_{\bX}$, but we proceed in this way to simplify.

Even if each $\bx_i$ is now considered as a realization of a random vector $\bX_i$, we study the asymptotic behaviour of the generalized posterior distribution under Tukey's biweight improper model as defined in \eqref{eq:postTukey}. This is sensible given that in a Bayesian model we can assume that $\bX_1, \ldots, \bX_n, \varepsilon_1, \ldots, \varepsilon_n$ and $\bbeta$ are all independent, which makes the posterior distribution independent of the assumed distribution of each $\bX_i$. This distribution thus do not appear in the posterior distribution. In the Bayesian model, we can assume that $\bX_i \sim \mu_{\bX}$. More details are provided in \autoref{sec:justification_pi_n}.

To study the asymptotic behaviour of the generalized posterior distribution under Tukey's biweight improper model, we apply Theorem 4 in \cite{miller2021asymptotic}. This is a generic result that allows to study the large-sample properties of a generic (possibly generalized) posterior distribution. Verifying the assumptions of this generic result requires considerable work. To use a similar notation as in that paper and to highlight a dependence on $n$, we rewrite the generalized posterior density in \eqref{eq:postTukey} as
\[
 \pi_n(\bbeta) := \frac{\pi(\bbeta) \exp(- n f_n(\bbeta))}{\int_{\re^p} \pi(\bbeta) \exp(- n f_n(\bbeta))  \, \d\bbeta}, \quad \bbeta \in \re^p,
\]
where $f_n: \re^p \rightarrow \re$ is a function such that
\[
 f_n(\bbeta) =w \log \hat\sigma_n + \frac{w}{n} \sum_{i=1}^n \varrho\left(\frac{y_i - \bx_i^T \bbeta}{\hat\sigma_n}\right), \quad \bbeta \in \re^p,
\]
with $\varrho = - \log g$, $-\log g$ being as in \eqref{eqn:biweight}, and $\hat\sigma_n$ being a generic robust estimator. We make the dependence of $\pi_n$ on the data implicit to simplify the notation. The function $\pi_n$ also depends on the tempering parameter $w > 0$ and $k > 0$ (the tuning parameter in the function $\varrho = -\log g$, recall \eqref{eqn:biweight}). These tuning parameters are considered fixed and known. Let us define $m_n := \int_{\re^p} \pi(\bbeta) \exp(- n f_n(\bbeta))  \, \d\bbeta$.

In order to state our large-sample result, we make assumptions on the components of $\pi_n$. The assumptions are similar to those in classic references (see, e.g., \cite{maronna1981asymptotic} and \cite{maronna2019robust}). We first present weak assumptions on $f_0$, the PDF of each error $\varepsilon_i$, which has an impact on the component $\by = (y_1, \ldots, y_n)^T$ of $\pi_n$.

   \begin{Assumption}\label{ass:error}
   The PDF $f_0$ is symmetric with respect to the origin, strictly decreasing from the origin and continuously differentiable. Also, $\E|\varepsilon| < \infty$,  $\varepsilon \sim f_0$.
 \end{Assumption}

 The main requirement is that $\E|\varepsilon| < \infty$,  $\varepsilon \sim f_0$. The other requirements are regularity conditions. \autoref{ass:error} is verified when $f_0$ is, for instance, the standard normal PDF or a Student's $t$ PDF, which can generate outliers. The PDF $f_0$ can additionally take the form of an $\epsilon$-contamination data generating process, that is $f_0 = (1 - \epsilon) f_1 + \epsilon f_2$ with $f_1$ the standard normal PDF and $\epsilon \in (0, 1)$, as long as $f_2$ is symmetric, strictly decreasing and continuously differentiable. The PDF $f_2$ can thus be a normal PDF with mean 0 and standard deviation $\sigma_\text{out.} > 1$ to generate outliers.

 We now present assumptions on $\mu_{\bX}$ which hold in great generality.

\begin{Assumption}\label{ass:mu_X}
  The distribution $\mu_{\bX}$ is such that the components in $(X_1, \ldots, X_p)^T := \bX \sim \mu_{\bX}$ are continuous random variables (except the first one) and we denote by $f_{\bX}$ the PDF of $\mu_{\bX}$. Also, $\E[|X_j|^3] < \infty$ for all $j$.
\end{Assumption}

To prove our theoretical result, we make use of a strongly consistent estimator $(\hat\bbeta_n, \hat\sigma_n)$ which converges to $(\bbeta_0, \sigma_0)$ as $n \rightarrow \infty$ with probability 1. The randomness here comes from the random variables $\bZ_1 = (\bX_1, Y_1), \ldots, \bZ_n = (\bX_n, Y_n)$, meaning that the convergence holds for almost all realizations $\{\bx_i, y_i\}_{i=1}^n$, or in other words, for all data sets that can be obtained under the true generating model. We thus continue with assumptions on this strongly consistent estimator $(\hat\bbeta_n, \hat\sigma_n)$. They either explicitly hold for Tukey’s biweight M-estimator $(\hat{\bbeta}_{\text{TM}}, \hat{\sigma}_{\text{TM}})$ or are expected to hold. In particular, under the assumptions of this sub-section, this estimator is strongly consistent \citep{maronna1981asymptotic}. Also, we essentially assume the following: outside of a compact subset of $\re^p$ that can be as large as we want and for large enough $n$,
\[
 f_n(\bbeta) - f_n(\hat\bbeta_n) = \frac{w}{n} \sum_{i=1}^n \varrho\left(\frac{y_i - \bx_i^T \bbeta}{\hat\sigma_n}\right) - \frac{w}{n} \sum_{i=1}^n \varrho\left(\frac{y_i - \bx_i^T \hat\bbeta_n}{\hat\sigma_n}\right) > 0,
\]
where $\hat\bbeta_n$ is defined as a zero of the derivative of $(1 / n) \sum_{i=1}^n \varrho((y_i - \bx_i^T \bbeta) / \hat\sigma_n)$.

  \begin{Assumption}\label{ass:estimator2} \,
  \begin{itemize}

  \item For almost all realizations $\{\bx_i, y_i\}_{i=1}^n$ (from the true generating model),

  \begin{itemize}

    \item $(\hat\bbeta_n, \hat\sigma_n) \rightarrow (\bbeta_0, \sigma_0)$ as $n \rightarrow \infty$,

    \item $f_n'(\hat\bbeta_n) = \mathbf{0}$ for all $n$, where $f_n'$ is the gradient of $f_n$,

    \item $\hat\sigma_n \in (0, \infty)$ and $\hat\beta_{nj} \in (-\infty, \infty)$ for all $n$ and all $j$,

    \item there exists a constant $C > 0$ (that can be chosen as large as we want) such that $\liminf_n \inf_{\bbeta \in B_C(\hat\bbeta_n)^\mathsf{c}} (f_n(\bbeta) - f_n(\hat\bbeta_n)) > 0$, where $B_C(\hat\bbeta_n) = \{\bbeta \in \re^p: \|\bbeta - \hat\bbeta_n\| < C\}$, $\| \, \cdot \|$ denoting the Euclidean norm.

  \end{itemize}

  \end{itemize}
 \end{Assumption}

 To prove our theoretical result, we also make use of the following lemma, which is interesting in its own right as it characterizes the large-sample behaviour of $f_n$.

 \begin{Lemma}\label{lemma:large-sample}
  Suppose Assumptions \ref{ass:error}-\ref{ass:estimator2} hold. Each following statements hold for almost all realizations $\{\bx_i, y_i\}_{i=1}^n$ (from the true generating model):
  \begin{description}
   \item[(a)] The sequence of functions $\{f_n\}$ is (uniformly) \textit{equicontinuous}\footnote{For a definition of equicontinuity, see, e.g., \cite{miller2021asymptotic}.}.
   \item[(b)] The sequence of functions $\{f_n\}$ converges pointwise: for any $\bbeta \in \re^p$, $f_n(\bbeta) \rightarrow \bar{f}(\bbeta)$ as $n \rightarrow \infty$, where
   \[
   \bar{f}(\bbeta) = w \log \sigma_0 + w \E\left[\varrho\left(\frac{Y - \bX^T \bbeta}{\sigma_0}\right)\right],
   \]
   with $\bX \sim \mu_{\bX}$ and $Y = \bX^T \bbeta_0 + \sigma_0 \varepsilon$, $\varepsilon \sim f_0$ and independent of $\bX$. As a consequence of Result (a), the convergence is uniform on any compact subset of $\re^p$.
   \item[(c)] The function $\bar{f}$ admits a unique global minimum at $\bbeta_0$.
   \item[(d)] For any $r > 0$, $\liminf_n \inf_{\bbeta \in B_r(\hat\bbeta_n)^\mathsf{c}} (f_n(\bbeta) - f_n(\hat\bbeta_n)) > 0$.
   \end{description}
 \end{Lemma}

 The function $\bar{f}$ in \autoref{lemma:large-sample} can be seen as the Kullback--Leibler (KL) divergence (up to an additive constant) of the true model from the assumed Tukey's biweight improper model, if we extend the definition of the KL divergence to include an improper distribution. More precisely,
    \[
     \bar{f}(\bbeta) = \E\left[\log \frac{p_0(Y \mid \bX) \,  f_{\bX}(\bX)}{p_{\bbeta}(Y \mid \bX) \,  f_{\bX}(\bX)}\right] + \textsf{cst}, \qquad \bbeta \in \re^p,
    \]
    where $p_0$ is the (true) conditional PDF of $Y$ given $\bX$, and
    \[
     p_{\bbeta}(y \mid \bx) = \left[\frac{1}{\sigma_0} g\left(\frac{y - \bx^T \bbeta}{\sigma_0}\right)\right]^w, \qquad y \in \re, \bx \in \re^p, \bbeta \in \re^p,
    \]
    which is viewed as the assumed improper ``conditional density function'', \textsf{cst} being a constant that does not depend on $\bbeta$. The KL divergence and its minimizer play a crucial role in establishing where the posterior mass concentrates for a proper (possibly misspecified) model and a Bernstein--von Mises result (see, e.g., \cite{kleijn2012bernstein}). Interestingly, this is mirrored in our analysis of an improper model. Note that, if $f_0$ is not symmetric nor monotonic, the minimizer of such a divergence may not be unique or may be different from $\bbeta_0$ \citep[Section 10.1]{maronna2019robust}.

    We finish with non-restrictive assumptions on the prior PDF $\pi$.

 \begin{Assumption}\label{ass:prior2}
  The prior PDF $\pi$ is continuous at $\bbeta_0$ and $\pi(\bbeta_0) > 0$.
 \end{Assumption}

    We are now ready to present the main result of this sub-section.

 \begin{Theorem}\label{Thm:large-sample}
Suppose Assumptions \ref{ass:error}-\ref{ass:prior2} hold. As $n \rightarrow \infty$, for almost all realizations $\{\bx_i, y_i\}_{i=1}^n$ (from the true generating model),
 \begin{description}
   \item[(a) ] the mass concentrates at $\bbeta_0$:
   \[
   \int_{B_r(\bbeta_0)} \pi_n(\bbeta) \, \d\bbeta \rightarrow 1 \quad \text{for all $r > 0$};
\]

     \item[(b)] a Laplace approximation holds: $m_n$ is asymptotically equivalent to
     \[
      \frac{\exp(-n f_n(\hat\bbeta_n)) \, \pi(\bbeta_0)}{|\det(\bH_0)|} \left(\frac{2\pi}{n}\right)^{p / 2},
     \]
     where
     \[
 \bH_0 = \frac{w}{\sigma_0^2} \E[\varrho''(\varepsilon)] \E[\bX \bX^T],
\]
a positive definite matrix;

\item[(c)] a Bernstein--von Mises result holds:
\[
 \int_{\re^p}\left|\pi_n(\bbeta) - \frac{1}{(2\pi)^{p/2}} \det(n \bH_0)^{1/2} \exp\left(-\frac{n}{2} (\bbeta - \hat{\bbeta}_n)^T \bH_0 (\bbeta - \hat{\bbeta}_n)\right) \right| \d\bbeta \rightarrow 0.
\]
\end{description}
\end{Theorem}

Results such as Result (a) are often referred to as \textit{consistency results} in the literature on large-sample properties of Bayesian methods as they indicate that, under some regularity conditions, estimators like MAP converge towards the true value. Here, such estimators are strongly consistent as the convergence holds for almost all realizations $\{\bx_i, y_i\}_{i=1}^n$. Result (b) describes the asymptotic behaviour of $m_n$ and Result (c) that of $\pi_n$. The latter thus asymptotically behaves like a concentrating normal distribution, centered at a consistent estimator $\hat\bbeta_n$ with a covariance matrix of $(n \bH_0)^{-1}$. The (asymptotic) covariance matrix characterizes the (asymptotic) uncertainty quantification.

We now use this characterization of the (asymptotic) uncertainty quantification to calibrate the generalized posterior distribution under Tukey's biweight improper model. We focus on the case where $f_0$ is the standard normal PDF as it allows to evaluate the large-sample properties of Tukey's biweight improper model and to calibrate the generalized posterior distribution in two important scenarios, both at once.

Firstly, it allows to evaluate its large-sample properties in the absence of outliers through a comparison with the gold-standard normal linear regression. Analyses of robust approaches in the traditional robustness literature typically include this scenario.

Secondly, it allows to evaluate its large-sample properties in the case where the outliers are far enough from the bulk of the data and the latter group is generated by the normal linear regression model. Indeed, we understand from \autoref{sec:robustness} that, in this scenario, the generalized posterior distribution under Tukey's biweight improper model essentially corresponds to that based only on the bulk of the data when using, as proposed, the estimator $\hat\sigma_{\text{TM}}$. In this scenario, $\{\bx_i, y_i\}_{i=1}^n$ on which the generalized posterior distribution is based represents the bulk of the data as the outliers, say $\{\bx_i, y_i\}_{i=n+1}^{n_{\text{Total}}}$, are excluded from the generalized posterior distribution. This scenario can be seen as corresponding to the $\epsilon$-contamination data generating process described above with $f_2$ a normal PDF with mean 0 and standard deviation $\sigma_\text{out.} \rightarrow \infty$. As mentioned in \autoref{sec:connection}, the bulk of the data is often expected and observed in practice (as in the examples of Sections \ref{sec:example1} and \ref{sec:example_reserve}) to be more closely aligned with the assumptions of the go-to light-tailed model, representing here the normal linear regression model. In this scenario, the Bayesian normal linear regression is well specified for the bulk of the data.

In both scenarios, we want to compare the asymptotic covariance matrix under Tukey's biweight improper model to that of the Bayesian normal linear regression model when estimated using $\{\bx_i, y_i\}_{i=1}^n$, which will allow to evaluate whether uncertainty quantification is meaningful under Tukey's biweight improper model. In \autoref{sec:info_estimation}, we present the posterior distribution under the normal model, which is seen to asymptotically behave like a concentrating normal distribution as well, but with a covariance matrix of $(\sigma_0^2 / n) \E[\bX \bX^T]^{-1}$ (this can be proved similarly as \autoref{Thm:large-sample}). The consistent estimator in this case can be OLS, which asymptotically also has a normal distribution with a covariance matrix of  $(\sigma_0^2 / n) \E[\bX \bX^T]^{-1}$. This implies that CIs of probability $1-\alpha$ (e.g., $1 - \alpha = 0.95$) under the normal model have frequentist coverage of $1 - \alpha$ (see, e.g., \cite{miller2021asymptotic}) and thus that uncertainty quantification is meaningful with this model. Therefore, a way to obtain meaningful uncertainty quantification with Tukey's biweight improper model is to match its asymptotic covariance matrix, given by $(n \bH_0)^{-1}$, with that under the normal model. \emph{This is achieved by setting $w = \E[\varrho''(\varepsilon)]^{-1}$, which is what we propose.} Let us say we do not and instead set $w = 1$, then $\E[\varrho''(\varepsilon)]^{-1} = 4.83$ if $k = 4.685$ and $f_0$ is the standard normal. This means that CIs are (asymptotically) larger than under the normal model by a factor of $\sqrt{4.83} = 2.20$. By selecting $w$ based on the proposed approach, the HPD CIs with Tukey's biweight improper model are similar to those obtained with the normal model based only on the bulk of the data, as observed in the examples of Sections \ref{sec:example1} and \ref{sec:example_reserve}. Note that the expectation $\E[\varrho''(\varepsilon)]$ can be numerically evaluated using adaptive quadrature.

There exist generic and data-driven approaches to calibrate the generalized posterior distribution (see, e.g., \cite{wu2023comparison} for a recent review and a comparison of popular approaches). The approach proposed here is grounded in the classic robustness literature by the structure assumed on the true generating model (see, e.g., \cite{maronna2019robust}). In this literature, it is typically assumed that the true generating model is the normal linear regression model (in the context of linear regression), but it is assumed that a part of the data set is possibly contaminated. An advantage of the proposed approach to calibrate the generalized posterior distribution is its simplicity.

\begin{figure}[ht]
 \centering\small

 $\begin{array}{ccc}
 \vspace{-1mm}
  \hspace{-2mm} \includegraphics[width=0.34\textwidth]{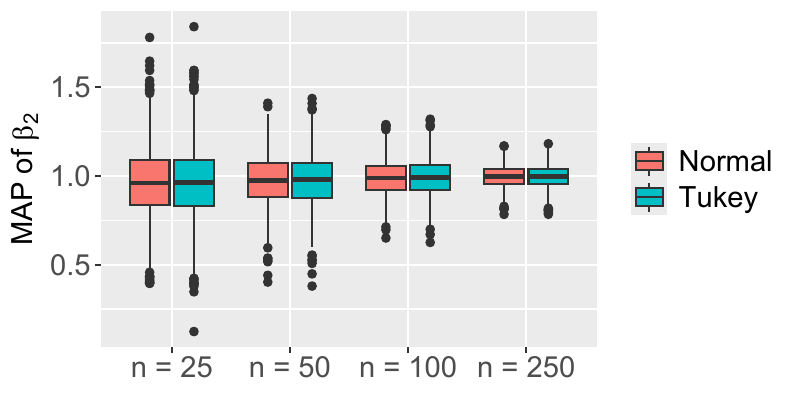} & \hspace{-4.5mm} \includegraphics[width=0.34\textwidth]{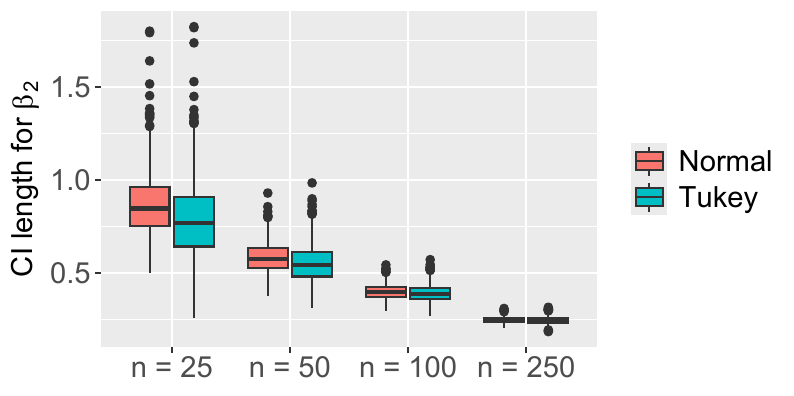} & \hspace{-4.5mm} \includegraphics[width=0.34\textwidth]{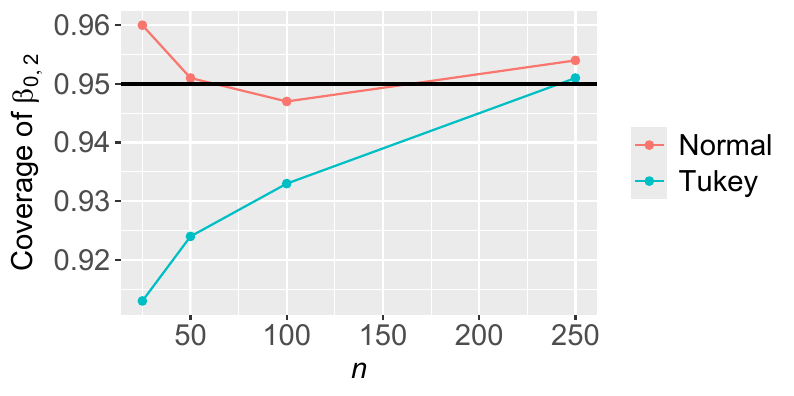} \cr
   \hspace{-8mm} \textbf{(a) } & \hspace{-12mm} \textbf{(b)} & \hspace{-9mm} \textbf{(c)}
  \end{array}$
  \vspace{-3mm}
\caption{Results of a Monte Carlo study with $p = 2$, $\bbeta_0 = (1, 1)^T$, $\sigma_0 = 1$, and $\mu_{\bX}$ and $f_0$ corresponding to the standard normal. For each value of $n$, 1,000 data sets are simulated, and for each data set, estimates are computed under Tukey's biweight improper model and the normal model. In (a), the MAP estimate of $\beta_2$ is shown. In (b), the HPD CI length for $\beta_2$ is shown. In (c), we present the coverage of the true coefficient value $\beta_{0, 2} = 1$ by the HPD CI.}\label{fig:large-sample}
\end{figure}

We finish this sub-section with results of a numerical experiment to illustrate \autoref{Thm:large-sample}. The experiment is a Monte Carlo study with $p = 2$, $\bbeta_0 = (1, 1)^T$, $\sigma_0 = 1$, and $\mu_{\bX}$ and $f_0$ both corresponding to the standard normal. We show results for increasing values of $n$, from $25$ to $250$. For each value of $n$, 1,000 data sets are simulated, and for each data set, estimates are computed, under both Tukey's biweight improper model and the normal model. The tuning parameters of the former are set as proposed in this section. The results are presented in \autoref{fig:large-sample}. They are consistent with the discussions of the previous paragraphs.

\normalsize

 \section{Second example: Reserve estimation}\label{sec:example_reserve}

In this section, we provide a second example where a significant difference is observed between estimation results obtained by a frequentist and a traditional Bayesian practitioner, the results of the latter being more influenced by the presence of outliers. The data set is that presented in \cite{VERRALL199175}, which takes the form of a \textit{development triangle} constructed from data provided in \cite{taylor1983second}. The data set is presented in \autoref{tab:triangle}. A development triangle is a way to present insurance claims by year the accidents occurred (the lines in the triangle) and the number of years between the accidents and the payments (the columns in the triangle). There may be time between accidents and payments because accidents are not reported immediately and there may be outstanding claims at the end of a given year.

\begin{table}[ht]
\footnotesize
 \centering
\begin{tabular}{l rrrrrrrrrr}
\toprule
\textbf{Accident} &  \multicolumn{10}{l}{\textbf{Number $j$ of years between the accidents}} \cr
\textbf{year $i$} &  \multicolumn{10}{l}{\textbf{and the payments}} \cr
\cline{2-11}
 & 0 & 1 & 2 & 3 & 4 & 5 & 6 & 7 & 8 & 9 \cr
\midrule
 0 & 357848 & 766940 & 610542 & 482940 & 527326 & 574398 & 146342 & 139950 & 227229 & 67948 \\
  1 & 352118 & 884021 & 933894 & 1183289 & 445745 & 320996 & 527804 & 266172 & 425046 &  \\
  2 & 290507 & 1001799 & 926219 & 1016654 & 750816 & 146923 & 495992 & 280405 &  &  \\
  3 & 310608 & 1108250 & 776189 & 1562400 & 272482 & 352053 & 206286 &  &  &  \\
  4 & 443160 & 693190 & 991983 & 769488 & 504851 & 470369 &  &  &  &  \\
  5 & 396132 & 937085 & 847498 & 805037 & 705960 &  &  &  &  &  \\
  6 & 440832 & 847631 & 1131398 & 1063269 &  &  &  &  &  &  \\
  7 & 359480 & 1061648 & 1443370 &  &  &  &  &  &  &  \\
  8 & 376686 & 986608 &  &  &  &  &  &  &  &  \\
  9 & 344014 &  &  &  &  &  &  &  &  &  \\
 \bottomrule
\end{tabular}
\caption{Development triangle based on the data in \cite{taylor1983second}.} \label{tab:triangle}
\end{table}
\normalsize

Let us consider for example that the first line in such a triangle corresponds to 2011 and that there are 10 lines. The number in the first line first column is regarding accidents that occurred in 2011 and corresponds to the amount paid for some of these accidents during the same year the accidents occurred, that is 2011. The number in the first line second column is again regarding accidents that occurred in 2011 but corresponds to the amount paid for some of these accidents during the year following the accident year, that is 2012. And so on. We will use \textsf{AY} to represent the variable \textit{accident year}, and \textsf{DY} (development year) to represent the variable \textit{number of years between the accidents and the payments}.

Let us say that we are at the end of 2020 in our example. Only a part of the triangle has been observed: the upper left part. To establish financial reserve for accidents that occurred but for which amounts will be paid in the future, insurance companies make predictions for the part of the triangle that is unobserved: the lower right part. Given that the data in the triangle are positive and right skewed, a linear regression is often assumed with the dependent variable being the log of the variable \textit{amount paid} and the explanatory variables being \textsf{AY} and \textsf{DY} \citep[Chapter 19]{frees2009regression}. In such a model, the explanatory variables are each considered categorical, with each value of  \textsf{AY} and \textsf{DY} being a category, to model non-linear relationships between the dependent variable and the explanatory variables. In the data set presented in \cite{VERRALL199175}, each explanatory variable has 10 categories, implying that the model has \emph{20} parameters (19 regression coefficients and an error scale parameter). The number of observations is \emph{55}.

Let us now discuss estimation results. The sum of the absolute differences between the coefficients estimated with Tukey's biweight M-estimator and those estimated with the Bayesian LPTN model is $0.79$. To evaluate the magnitude of this difference, the same sum, but calculated between the coefficients estimated with Tukey's biweight M-estimator and those estimated with OLS, is $1.09$. The estimation results for the LPTN model were produced with $\rho = 0.88$, which is the value that maximizes the posterior density and is among the values that produce coefficient estimates that are the closest to Tukey's biweight M-estimates. The results with the Student's $t$ model are similar to those with the LPTN model, but are more influenced by the presence of outliers.

\begin{figure}[ht]
  \centering\small
  $\begin{array}{cc}
\vspace{-2mm}\hspace{-3mm}  \includegraphics[width=0.5\textwidth]{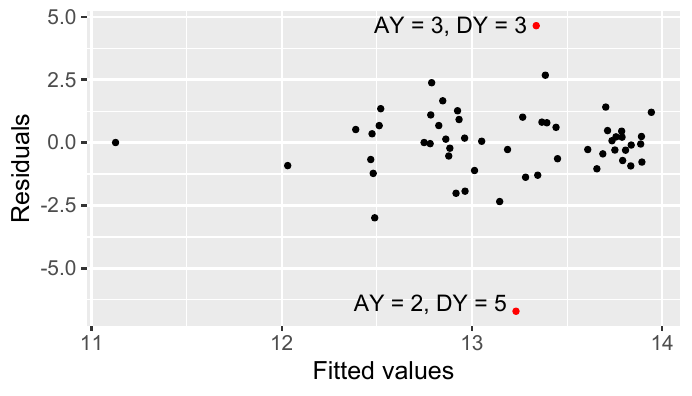} & \hspace{-3mm} \includegraphics[width=0.5\textwidth]{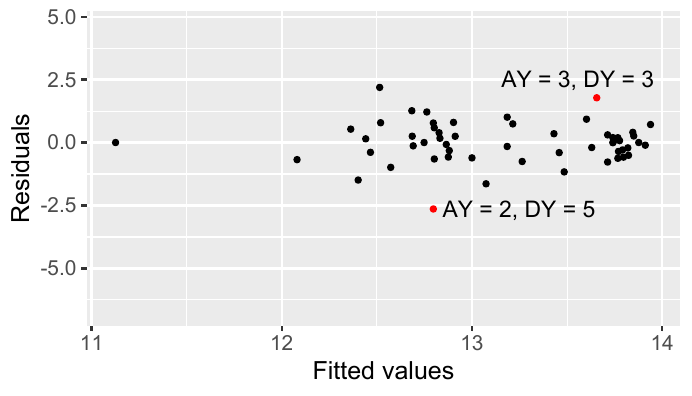} \cr
 \vspace{2mm}\textbf{(a) Tukey's biweight M-estimation} & \hspace{-5mm} \textbf{(b) OLS estimation} \cr
 \vspace{-2mm}\hspace{-3mm}  \includegraphics[width=0.5\textwidth]{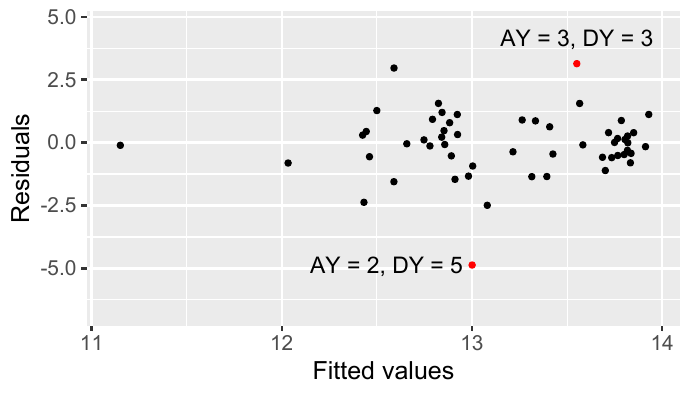} &  \hspace{-3mm}  \includegraphics[width=0.5\textwidth]{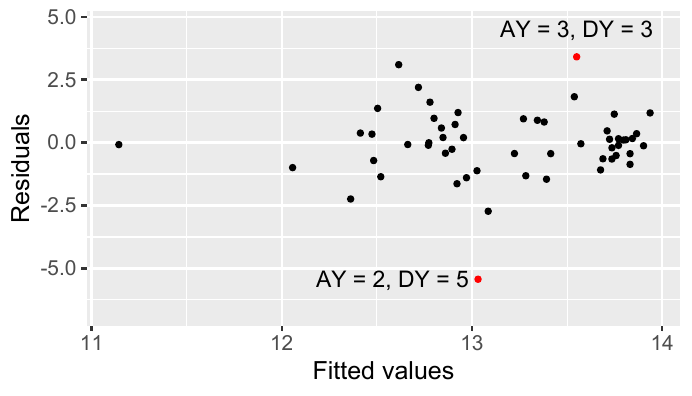}\cr
    \textbf{(c) Bayesian LPTN estimation}  & \hspace{-5mm}   \textbf{(d) Bayesian Student's $t$ estimation}
  \end{array}$\vspace{-2mm}
  \caption{Standardized residuals against fitted values computed from (a) Tukey's biweight estimates, (b) OLS estimates, (c) Bayesian LPTN estimates and (d) Bayesian Student's $t$ estimates.}\label{fig:residuals}
 \end{figure}

We understood that the difference between Tukey's biweight M-estimation and the Bayesian LPTN estimation is due to more pronounced influence of the outliers on the latter. Indeed, we identified two outliers and, for each approach, we compare the estimation results obtained with and without the outliers. There is a slight difference with Tukey's biweight M-estimation but, with Bayesian LPTN estimation, the difference is significant. Also, Tukey's biweight M-estimation is the closest to OLS estimation based on the data set without the outliers. The outlier detection was based on an analysis of the standardized residuals computed with Tukey's biweight M-estimates. These standardized residuals are presented in \autoref{fig:residuals} against fitted values, along with the versions computed with the OLS, Bayesian LPTN and Student's $t$  estimates.

\normalsize

 In \autoref{fig:residuals}, we observe that, with Tukey's biweight M-estimates, the standardized residuals of the non-outlying observations appear overall better dispersed than in the three other cases. Between \autoref{fig:residuals} (a) and \autoref{fig:residuals} (c), we observe a significant difference in the residuals of the observation with $\textsf{AY} = 2$ and $\textsf{DY} = 5$. The absolute difference between the coefficient estimates associated with $\textsf{DY} = 5$ is the largest. The exponential of the estimated coefficient is $1.31$ with Tukey's biweight M-estimation and $1.09$ with the Bayesian LPTN estimation, indicating a significant difference in terms of predictions for the variable \textit{amount paid} (recall that the dependent variable in the model is a log transformation of this variable).

 Under the generalized Bayesian framework, Tukey's biweight improper model studied in \autoref{sec:Tukeymodel} leads to essentially the same estimation as its M-estimator counterpart. For instance, the MAP estimate of the coefficient associated with $\textsf{DY} = 5$ is $0.27$ whose exponential is $1.31$. The 95\% HPD CI for this coefficient is $(-0.18, 0.61)$. This MAP estimate and CI are similar to those obtained with the normal model based on the data set excluding the outliers. Adopting the generalized Bayesian framework thus allows the Bayesian practitioner to obtain similar results as the frequentist practitioner.

 \section{Conclusion}\label{sec:conclusion}

 In this paper, we investigated the connection between classical Bayesian and frequentist approaches to robustness against outliers. The classical Bayesian approach is to adapt the model to the presence of outliers by using a heavy-tailed distribution. On the frequentist side, the classical approach is to derive a robust estimator through a modification of the log-likelihood function or its derivative. In many situations, both approaches can be connected; it is for instance the case when a M-estimator can be viewed as the maximum likelihood estimator of a proper heavy-tailed model. As traditional Bayesians require the heavy-tailed model to be proper, establishing the connection is however not possible in the situations where the modified log-likelihood function (or its derivative) corresponds to an improper heavy-tailed model. There thus exists a fundamental difference between classical Bayesian and frequentist approaches to robustness against outliers. A consequence is that a traditional Bayesian practitioner does not have access to the same range of tools as a frequentist practitioner. In the paper, we demonstrated through two real-data examples that this difference may lead to significantly different estimation results for the traditional Bayesian practitioner, with results that are more influenced by the presence of outliers. The adoption of the generalized Bayesian framework by the Bayesian practitioner allows this practitioner to have access to the same range of tools, thus representing a way to completely reconcile the Bayesian and frequentist approaches to robustness.

  In the paper, we focused on linear regression for the explanations and examples. As Tukey's biweight M-estimator is a popular robust estimator in linear regression corresponding to an improper model, it is important to study its generalized Bayesian counterpart. A contribution of this paper is to provide an extensive theoretical study and, more precisely, a characterization of its robustness and large-sample behaviour. In particular, a Bernstein--von Mises result has been established, allowing to calibrate the generalized posterior distribution for meaningful uncertainty quantification.

  By focusing on the classical approaches to robustness and linear regression, we did not broadly discuss the different robust approaches. In Bayesian linear regression, other approaches have been proposed such as using a mixture distribution \citep{box1968bayesian, hamura2020log, HAMURA2024110130}. There also exist generic approaches, meaning approaches that can be applied to gain robustness in a general modelling framework. \cite{hooker2014bayesian} proposed a methodology using disparities and kernel density estimators. \cite{ghosh2016robust} proposed an approach which instead consists of using a density power divergence. The latter is discussed more broadly in the context of divergence-based loss functions in \cite{jewson2018principles} and extended to accommodate intractable likelihood functions in \cite{matsubara2022robust}. \cite{wang2017robust} proposed a methodology based on reweighing the likelihood terms of the observations depending on their alignment with the general trend. A related approach is that of \cite{miller2019robust} based on  coarsening and divergence measures which corresponds to raising the likelihood function to a power when considering the Kullback--Leibler divergence. \cite{wang2018robust} use localization and empirical Bayes to propose a generic approach. \cite{lewis2021bayesian} propose to condition on statistics rather than the whole data set to avoid inference contamination. Interestingly, Tukey's biweight M-estimator is used as a statistic, which can be seen as a different way of incorporating this estimator in Bayesian analysis. Recently, \cite{bhatia2024bayesian} proposed a different generic approach which is instead based on a robust MCMC scheme. With this approach, the robustness comes from the algorithm which is used for inference. The approaches of \cite{ghosh2016robust}, \cite{wang2017robust}, \cite{jewson2018principles}, \cite{miller2019robust} (at least when considering the Kullback--Leibler divergence) and \cite{matsubara2022robust} all fit within the generalized Bayesian framework of \cite{bissiri2016general}. Some of the approaches mentioned in this paragraph do not have a clear connection with frequentist robust approaches, like that of \cite{bhatia2024bayesian}. In future work, it would be interesting to investigate if a parallel can be made with frequentist robust approaches.

\bibliographystyle{rss}
\bibliography{references}

\appendix

\section{About M-estimators and Bayesian models}\label{sec:info_M_Bayes}

In this section, we first present the formal definition of the weight function, which will be in particular derived for the M-estimators and the Bayesian models on which we focused in the paper. It is defined as follows: $W(\varepsilon) := \psi(\varepsilon) / \varepsilon$ when $\varepsilon \neq 0$ and $W(\varepsilon) = \psi'(0)$ when $\varepsilon = 0$, with $\psi$ being the derivative of $-\log g$ (up to a multiplicative constant); see Section 2.2 of \cite{maronna2019robust}. The derivative of the function in \eqref{eqn:logL} involves terms like $\psi((y_i - \bx_i^T \bbeta) / \sigma)$. If $f$ was the standard normal PDF, $\psi$ would be the identity function. To view an M-estimator as weighted OLS, we write
\[
 \psi\left(\frac{y_i - \bx_i^T \bbeta}{\sigma}\right) = \frac{y_i - \bx_i^T \bbeta}{\sigma} \frac{\psi\left(\frac{y_i - \bx_i^T \bbeta}{\sigma}\right)}{\frac{y_i - \bx_i^T \bbeta}{\sigma}} = \frac{y_i - \bx_i^T \bbeta}{\sigma} \, W\left(\frac{y_i - \bx_i^T \bbeta}{\sigma}\right),
\]
when $(y_i - \bx_i^T \bbeta) / \sigma \neq 0$. The weight function $W$ thus corresponds to the weight assigned to a standardized residual $(y_i - \bx_i^T \bbeta) / \sigma$ (or, equivalently, to a data point) in the estimating equation, that is a zero of the derivative of the function in \eqref{eqn:logL}.

\subsection{Huber M-estimator}

For the Huber M-estimator, we have that
\begin{align*} % \label{eqn:huber}
\varrho(\varepsilon) := -2\log g(\varepsilon)=\left\{
\begin{array}{lcc}
                                                      \varepsilon^2  & \text{ if } & |\varepsilon|\leq k, \cr
                                                     2 k|\varepsilon| - k^2  & \text{ if } & |\varepsilon|>k, \cr
\end{array}
\right.
\end{align*}
\begin{align*} % \label{eqn:huber}
\psi(\varepsilon) = \varrho'(\varepsilon) / 2 =\left\{
\begin{array}{lcc}
                                                      \varepsilon  & \text{ if } & |\varepsilon|\leq k, \cr
                                                      k\,\sgn(\varepsilon)  & \text{ if } & |\varepsilon|>k, \cr
\end{array}
\right.
\end{align*}
where $\sgn(\, \cdot \,)$ is the sign function, and
\[
 W(\varepsilon) = \min\{1, k / |\varepsilon|\}.
\]

\begin{figure}[ht]
 \centering\small
    \includegraphics[width=0.5\textwidth]{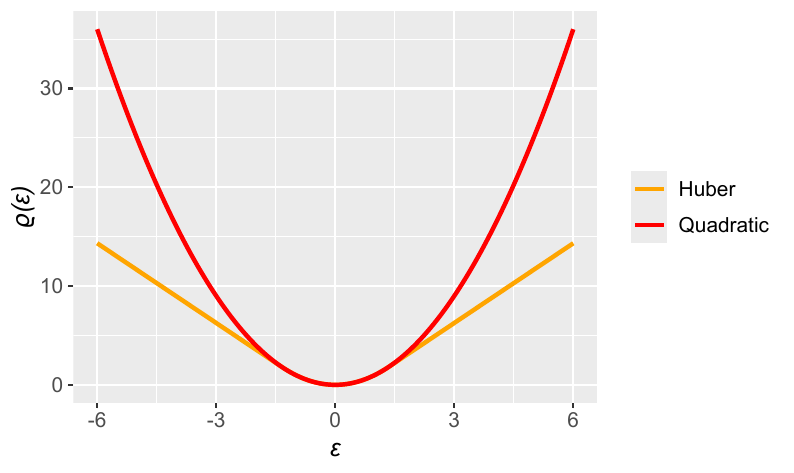}
  \vspace{-3mm}
\caption{$\varrho$ as a function of $\varepsilon$ when $\varrho$ is the quadratic function and the function associated to the Huber M-estimator with $k = 1.345$.}\label{fig:Hubervsquad}
\end{figure}

In \autoref{fig:Hubervsquad}, we present a comparison between the quadratic function (corresponding to the normal model) and the function $\varrho$ associated to the Huber M-estimator.
\normalsize

\subsection{Tukey's biweight M-estimator}\label{sec:biweight}

For Tukey's biweight M-estimator, we have that
\begin{align*}
\varrho(\varepsilon) := -\log g(\varepsilon)=\left\{
\begin{array}{lcc}
                                                      1-(1-(\varepsilon/k)^2)^3  & \text{ if } & |\varepsilon|\leq k, \cr
                                                      1 & \text{ if } & |\varepsilon|>k,
\end{array}
\right.
\end{align*}
\begin{align*} % \label{eqn:huber}
\psi(\varepsilon) = k^2 \varrho'(\varepsilon) / 6 = \varepsilon\left(1 - (\varepsilon / k)^2\right)^2 \ind(|\varepsilon| \leq k),
\end{align*}
and
\[
 W(\varepsilon) = \left(1 - (\varepsilon / k)^2\right)^2 \ind(|\varepsilon| \leq k).
\]
Note that
\[
 \varrho''(\varepsilon) = \frac{6}{k^2} \left[\left(1 - \left(\frac{\varepsilon}{k}\right)^2\right)^2 - 4 \left(\frac{\varepsilon}{k}\right)^2 \left(1 - \left(\frac{\varepsilon}{k}\right)^2\right)\right] \ind(|\varepsilon| \leq k),
\]
\[
 \varrho'''(\varepsilon) = -\frac{24\varepsilon}{k^4} \left[3\left(1 - \left(\frac{\varepsilon}{k}\right)^2\right) - 2 \left(\frac{\varepsilon}{k}\right)^2\right] \ind(|\varepsilon| \leq k).
\]
The first and second derivatives $\varrho'$ and $\varrho''$ are, like $\varrho$, continuous and bounded. The third derivatives $\varrho'''$ is bounded but not continuous (it is not continuous at $-k$ and $k$).

\begin{figure}[ht]
 \centering\small
 $\begin{array}{cc}
   \includegraphics[width=0.5\textwidth]{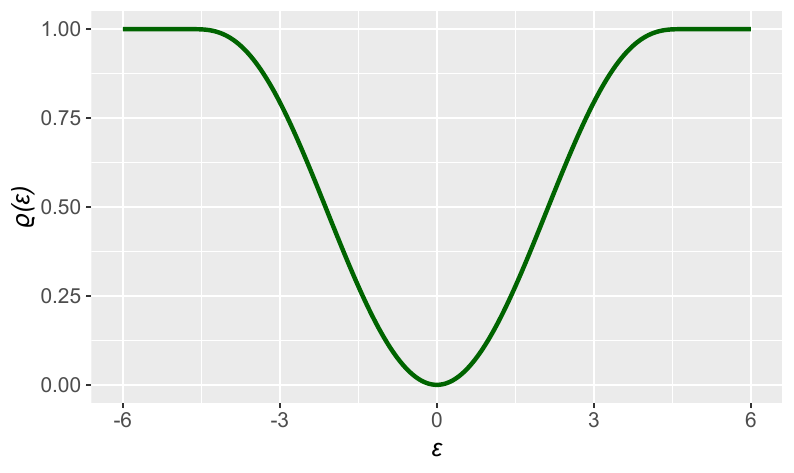} & \hspace{-5mm} \includegraphics[width=0.5\textwidth]{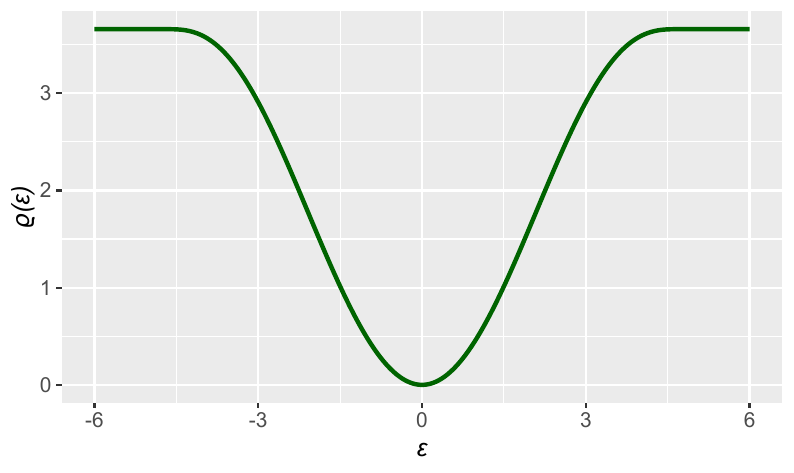} \cr
   \hspace{-0mm} \textbf{(a) } & \hspace{-5mm} \textbf{(b)}
  \end{array}$
  \vspace{-3mm}
\caption{(a) $\varrho$ as a function of $\varepsilon$ when $\varrho$ is the function associated to Tukey's biweight M-estimator with $k = 4.685$. (b) $\varrho$ as a function of $\varepsilon$ when $\varrho$ is the function associated to Tukey's biweight M-estimator defined through the alternative loss function in \eqref{eqn:biweight2} with $k = 4.685$.}\label{fig:biweight}
\end{figure}

In \autoref{fig:biweight} (a), we present the function $\varrho$ associated to Tukey's biweight M-estimator. In \autoref{fig:biweight} (b), we present the $\varrho$ function associated to Tukey's biweight M-estimator, but that defined through the alternative loss function in \eqref{eqn:biweight2}. In this figure, we do not present the quadratic function because these functions operate on a different scale.

\normalsize

As mentioned in \autoref{sec:M-estimators}, the M-estimator can be interpreted as the maximum likelihood estimator of an heavy-tailed model, but in this case the model is improper and only the function $g$ can be identified:
\begin{align*}
g(\varepsilon)=\left\{
\begin{array}{lcc}
                                                      \exp\left(-1+(1-(\varepsilon/k)^2)^3\right)  & \text{ if } & |\varepsilon|\leq k, \cr
                                                      \exp(-1) & \text{ if } & |\varepsilon|>k.
\end{array}
\right.
\end{align*}

\subsection{Student's $t$ model}

For the Student's $t$ model, we have that
\[
 f(\varepsilon) = \frac{\Gamma\left(\frac{\nu + 1}{2}\right)}{\sqrt{\pi \nu} \, \Gamma\left(\frac{\nu}{2}\right)} \left(1 + \frac{\varepsilon^2}{\nu}\right)^{- \frac{\nu + 1}{2}} = \frac{1}{m} \left(1 + \frac{\varepsilon^2}{\nu}\right)^{- \frac{\nu + 1}{2}}, \quad \varepsilon \in \re,
\]
where $\Gamma$ is the gamma function and $\nu > 0$ represents the degrees of freedom.

We can view maximum likelihood estimation of the model as M-estimation of the normal model. From this perspective, we have that
\[
 \varrho(\varepsilon) := -2\log g(\varepsilon) = (\nu + 1) \log\left(1 + \varepsilon^2 / \nu\right),
\]
\begin{align*} % \label{eqn:huber}
\psi(\varepsilon) =\frac{\nu}{2(\nu + 1)} \, \varrho'(\varepsilon) =\frac{\varepsilon}{1 + \varepsilon^2 / \nu},
\end{align*}
and
\[
 W(\varepsilon) = \frac{1}{1 + \varepsilon^2 / \nu}.
\]

\begin{figure}[ht]
 \centering\small
    \includegraphics[width=0.5\textwidth]{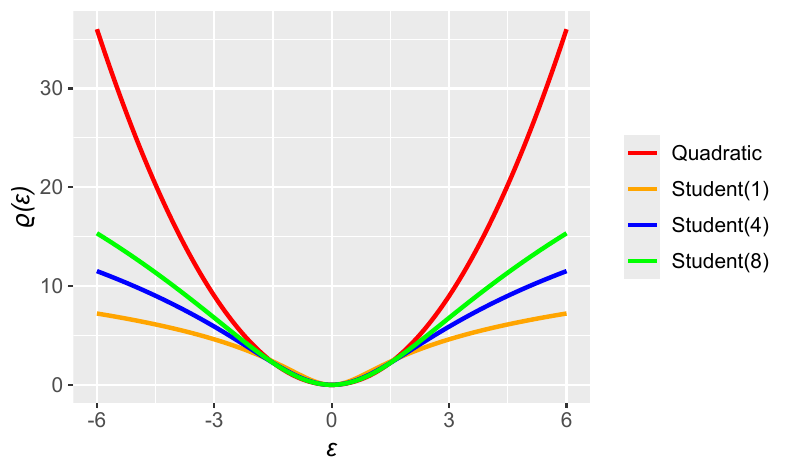}
  \vspace{-3mm}
\caption{$\varrho$ as a function of $\varepsilon$ when $\varrho$ is the quadratic function and the function associated to the Student's $t$ model with $\nu = 1$, $\nu = 4$ and $\nu = 8$.}\label{fig:Studentvsquad}
\end{figure}

In \autoref{fig:Studentvsquad}, we present a comparison between the quadratic function and the function $\varrho$ associated to the Student's $t$ model.
\normalsize

\subsection{LPTN model}

For the LPTN model, we have that
\begin{align*}
f(\varepsilon)=\left\{
\begin{array}{lcc}
                                                      \varphi(\varepsilon)  & \text{ if } & |\varepsilon|\leq \tau, \cr
                                                      \varphi(\tau)\,\frac{\tau}{|\varepsilon|}\left(\frac{\log \tau}{\log |\varepsilon|}\right)^{\lambda} & \text{ if } & |\varepsilon|>\tau.
\end{array}
\right.
\end{align*}
See \autoref{sec:heavy-tailed} for the details.

We can view maximum likelihood estimation of the model as M-estimation of the normal model. From this perspective, we have that
\begin{align*}
 \varrho(\varepsilon) &= -2(\log g(\varepsilon) - \log m + (1 / 2) \log(2\pi)) \cr
 &=-2(\log f(\varepsilon) + (1 / 2) \log(2\pi)) \cr
 & = \left\{
\begin{array}{lcc}
                                                      \varepsilon^2  & \text{ if } & |\varepsilon|\leq \tau, \cr
                                                      \tau^2 - 2\log(\tau) + 2\log|\varepsilon|-2\lambda \log(\log \tau) + 2\lambda \log(\log |\varepsilon|) & \text{ if } & |\varepsilon|>\tau,
\end{array}
\right.
\end{align*}
\begin{align*} % \label{eqn:huber}
\psi(\varepsilon) = \varrho'(\varepsilon) / 2  = \left\{ \begin{array}{lcc}
                                                      \varepsilon  & \text{ if } & |\varepsilon|\leq \tau, \cr
                                                         \frac{1}{\varepsilon}+ \frac{\lambda}{\varepsilon  \log|\varepsilon|} & \text{ if } & |\varepsilon|>\tau,
\end{array}
\right.
\end{align*}
and
\[
 W(\varepsilon) = \left\{ \begin{array}{lcc}
                                                      1  & \text{ if } & |\varepsilon|\leq \tau, \cr
                                                         \frac{1}{\varepsilon^2}+ \frac{\lambda}{\varepsilon^2 \log|\varepsilon|} & \text{ if } & |\varepsilon|>\tau.
\end{array}
\right.
\]

\begin{figure}[ht]
 \centering\small
    \includegraphics[width=0.5\textwidth]{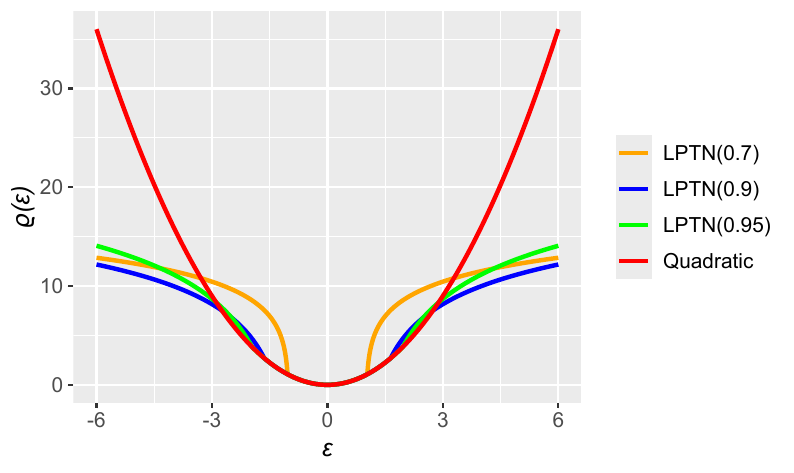}
  \vspace{-3mm}
\caption{$\varrho$ as a function of $\varepsilon$ when $\varrho$ is the quadratic function and the function associated to the LPTN model with $\rho = 0.7$, $\rho = 0.9$ and $\rho = 0.95$.}\label{fig:LPTNvsquad}
\end{figure}

In \autoref{fig:LPTNvsquad}, we present a comparison between the quadratic function and the function $\varrho$ associated to the LPTN model.
\normalsize

\section{Proofs of the results in \autoref{sec:Tukeymodel} and auxiliary results}\label{sec:proofs}

In this section, we present the proofs of the results in \autoref{sec:Tukeymodel}, in the same order as the results appeared. We also present Lemmas \ref{lemma1}-\ref{lemma:positive}. \autoref{lemma1} is used in the proof of \autoref{lemma:large-sample}, while Lemmas \ref{lemma:varrhopprim}-\ref{lemma:positive} are used in the proof of \autoref{Thm:large-sample}.

\begin{proof}[Proof of \autoref{prop:proper}]
 It is assumed that the prior distribution $\pi$ is proper and that $\hat{\sigma}_{\text{TM}} \in (0, \infty)$ for the data set $\{\bx_i, y_i\}_{i=1}^n$ at hand. For all $n \in \na$ and $w > 0$, we first prove that
 \[
  \int_{\re^p} \pi(\bbeta) \left[\prod_{i=1}^n \frac{1}{\hat{\sigma}_{\text{TM}}} g\left(\frac{y_i - \bx_i^T \bbeta}{\hat{\sigma}_{\text{TM}}}\right)\right]^w \d\bbeta < \infty,
 \]
 which implies that $\pi(\, \cdot \mid \by)$ defined in \eqref{eq:postTukey} is proper. Given that $g \leq 1$ and the prior distribution is proper,
 \begin{align*}
  \int_{\re^p} \pi(\bbeta) \left[\prod_{i=1}^n \frac{1}{\hat{\sigma}_{\text{TM}}} g\left(\frac{y_i - \bx_i^T \bbeta}{\hat{\sigma}_{\text{TM}}}\right)\right]^w \d\bbeta \leq \frac{1}{\hat{\sigma}_{\text{TM}}^{n w}} \int_{\re^p} \pi(\bbeta) \, \d\bbeta = \frac{1}{\hat{\sigma}_{\text{TM}}^{n w}} < \infty,
 \end{align*}
 for all $n \in \na$ and $w > 0$. Using the same arguments, we have that the moments of order $\kappa \in \re$ exist, provided that these moments exist under the prior distribution.
\end{proof}

\begin{proof}[Proof of \autoref{Thm:robustness}]
 We start with the proof of Result (c) (assuming Result (b)). Next, we prove Result (b) (assuming Result (a)). Finally, we provide the proof of Result (a), which is slightly longer.

 Result (c) is a direct consequence of Result (b) by Scheffé's lemma, which states that the convergence almost everywhere of a PDF is sufficient to ensure the convergence in distribution (see \cite{scheffe1947useful}). To prove Result (b), we rewrite $\pi_\vartheta(\bbeta \mid \by)$ for fixed $\bbeta \in A$ in order to exploit Result (a) and the result in \eqref{eqn:limitTukey}:
 \begin{align*}
 \pi_\vartheta(\bbeta \mid \by) &= \pi(\bbeta \mid \by_{\Oset^\mathsf{c}}) \, \left[\frac{\bar\sigma}{\hat\sigma}_\vartheta\right]^{w |\Oset^\mathsf{c}|} \frac{m(\by_{\Oset^\mathsf{c}}) \, \hat\sigma_\vartheta^{-w|\Oset|} \exp(-w|\Oset|)}{m_\vartheta(\by)} \cr
 &\hspace{40mm} \times \left[\prod_{i \in \Oset } \frac{g((y_i(\vartheta)-\bx_i(\vartheta)^T\bbeta)/\hat{\sigma}_\vartheta)}{\exp(-1)}\right]^w.
\end{align*}
Under Assumptions \ref{ass:prior} and \ref{ass:estimator}, we know by \autoref{prop:proper} that $0 < m(\mathbf{y}_{\Oset^\mathsf{c}}) < \infty$ and $0 < m_\vartheta(\by) < \infty$ for all $\vartheta$. For any $\bbeta \in A$,
\[
  \left[\frac{\bar\sigma}{\hat\sigma}_\vartheta\right]^{w |\Oset^\mathsf{c}|} \frac{m(\by_{\Oset^\mathsf{c}}) \, \hat\sigma_\vartheta^{-w|\Oset|} \exp(-w|\Oset|)}{m_\vartheta(\by)} \left[\prod_{i \in \Oset } \frac{g((y_i(\vartheta)-\bx_i(\vartheta)^T\bbeta)/\hat{\sigma}_\vartheta)}{\exp(-1)}\right]^w \rightarrow 1,
\]
as $\vartheta \rightarrow \infty$, by Result (a) and the result in \eqref{eqn:limitTukey} (under Assumptions \ref{ass:prior}-\ref{ass:estimator}). In fact, what is used is
\[
 \frac{g\left(\frac{y_i(\vartheta)-\bx_i(\vartheta)^T\bbeta}{\hat{\sigma}_\vartheta}\right)}{\exp(-1)} = \frac{g\left(\frac{y_i(\vartheta)-\bx_i(\vartheta)^T\bbeta}{\bar\sigma}\frac{\bar\sigma}{\hat{\sigma}_\vartheta}\right)}{\exp(-1)} \rightarrow 1,
\]
for all $i \in \Oset$ and $\bbeta \in A$ as $\vartheta \rightarrow \infty$ (under Assumptions \ref{ass:outliers}-\ref{ass:estimator} and using that $g$ is continuous and its definition), which is closely related to the result in \eqref{eqn:limitTukey}.

We now prove Result (a) using Lebesgue's dominated convergence theorem. We have that
\begin{align}\label{eqn:integral}
 &\frac{m_\vartheta(\by)}{\hat\sigma_\vartheta^{-w|\Oset|} \exp(-w|\Oset|)}\cr
  &= \int_{\re^p}  \pi(\bbeta) \left[\prod_{i \in \Oset^\mathsf{c}} \frac{1}{\hat{\sigma}_\vartheta} g\left(\frac{y_i - \bx_i^T \bbeta}{\hat\sigma_\vartheta}\right)\right]^w \left[\prod_{i \in \Oset} \frac{g((y_i(\vartheta) - \bx_i(\vartheta)^T \bbeta) / \hat\sigma_\vartheta)}{\exp(-1)}\right]^w \d\bbeta.
\end{align}
For any $\bbeta \in A$, under Assumptions \ref{ass:prior}-\ref{ass:estimator} and using that $g$ is continuous and its definition (as above),
\begin{align*}
 &\lim_{\vartheta \rightarrow \infty} \pi(\bbeta) \left[\prod_{i \in \Oset^\mathsf{c}} \frac{1}{\hat{\sigma}_\vartheta} g\left(\frac{y_i - \bx_i^T \bbeta}{\hat\sigma_\vartheta}\right)\right]^w \left[\prod_{i \in \Oset} \frac{g((y_i(\vartheta) - \bx_i(\vartheta)^T \bbeta) / \hat\sigma_\vartheta)}{\exp(-1)}\right]^w \cr
 &\qquad = \pi(\bbeta) \left[\prod_{i \in \Oset^\mathsf{c}} \frac{1}{\bar{\sigma}} g\left(\frac{y_i - \bx_i^T \bbeta}{\bar\sigma}\right)\right]^w,
\end{align*}
which corresponds to the integrand in $m(\by_{\Oset^\mathsf{c}})$.

To apply Lebesgue's dominated convergence theorem and thus obtain the result, there remains to prove that the integrand in \eqref{eqn:integral} is bounded above by a function which is independent of $\vartheta$ and integrable. Using that $g \leq 1$,
\begin{align*}
& \pi(\bbeta) \left[\prod_{i \in \Oset^\mathsf{c}} \frac{1}{\hat{\sigma}_\vartheta} g\left(\frac{y_i - \bx_i^T \bbeta}{\hat\sigma_\vartheta}\right)\right]^w \left[\prod_{i \in \Oset} \frac{g((y_i(\vartheta) - \bx_i(\vartheta)^T \bbeta) / \hat\sigma_\vartheta)}{\exp(-1)}\right]^w \cr
&\qquad \leq \pi(\bbeta) \left[\prod_{i \in \Oset^\mathsf{c}} \frac{1}{\hat\sigma_\vartheta} \right]^w \left[\prod_{i \in \Oset} \frac{1}{\exp(-1)}\right]^w.
\end{align*}
Under \autoref{ass:estimator}, $\hat\sigma_\vartheta \in (0, \infty)$ for all $\vartheta$ and $\hat\sigma_\vartheta \rightarrow \bar\sigma$ as $\vartheta \rightarrow \infty$, where $\bar\sigma \in (0, \infty)$. Therefore, there exists a constant $C > 0$ such that
\[
 \pi(\bbeta) \left[\prod_{i \in \Oset^\mathsf{c}} \frac{1}{\hat\sigma_\vartheta} \right]^w \left[\prod_{i \in \Oset} \frac{1}{\exp(-1)}\right]^w \leq C \, \pi(\bbeta),
\]
which is independent of $\vartheta$ and integrable. Therefore, by Lebesgue's dominated convergence theorem,
\[
\lim_{\vartheta \rightarrow \infty} \frac{m_\vartheta(\by)}{\hat\sigma_\vartheta^{-w|\Oset|} \exp(-w|\Oset|)} = m(\by_{\Oset^\mathsf{c}}).
\]
This concludes the proof. Note that it is possible to prove that $\E[\beta_j^k \mid \by] \rightarrow \E[\beta_j^k \mid \by_{\Oset^\mathsf{c}}]$ as long as $\E[\beta_j^k]$ exists under the prior distribution. Indeed,
\begin{align*}
 &\int_{\re^p} \beta_j^k \, \pi_\vartheta(\bbeta \mid \by) \, \d\bbeta \cr
  &=  \int_{\re^p} \beta_j^k \, \pi(\bbeta \mid \by_{\Oset^\mathsf{c}}) \, \left[\frac{\bar\sigma}{\hat\sigma}_\vartheta\right]^{w |\Oset^\mathsf{c}|} \frac{m(\by_{\Oset^\mathsf{c}}) \, \hat\sigma_\vartheta^{-w|\Oset|} \exp(-w|\Oset|)}{m_\vartheta(\by)} \left[\prod_{i \in \Oset } \frac{g\left(\frac{y_i(\vartheta)-\bx_i(\vartheta)^T\bbeta}{\hat{\sigma}_\vartheta}\right)}{\exp(-1)}\right]^w \d\bbeta,
\end{align*}
and
\[
 \left[\frac{\bar\sigma}{\hat\sigma}_\vartheta\right]^{w |\Oset^\mathsf{c}|} \frac{m(\by_{\Oset^\mathsf{c}}) \, \hat\sigma_\vartheta^{-w|\Oset|} \exp(-w|\Oset|)}{m_\vartheta(\by)} \left[\prod_{i \in \Oset } \frac{g((y_i(\vartheta)-\bx_i(\vartheta)^T\bbeta)/\hat{\sigma}_\vartheta)}{\exp(-1)}\right]^w
\]
converges to 1, as explained in the proof of Result (b) above, and it is bounded. Therefore, there exists a constant $C > 0$ such that
\[
 \beta_j^k \, \pi_\vartheta(\bbeta \mid \by) \leq C \, \beta_j^k \, \pi(\bbeta \mid \by_{\Oset^\mathsf{c}}),
\]
which is independent of $\vartheta$ and integrable (when $\E[\beta_j^k]$ exists under the prior distribution, which can be deduced using the same strategy as in the proof of \autoref{prop:proper}). Therefore, we obtain the result by Lebesgue's dominated convergence theorem.
\end{proof}

\begin{proof}[Proof of \autoref{lemma:large-sample}]
  To prove Result (a), we show that $f_n$ is Lipschitz continuous with the same constant for all $n$. We have that
  \begin{align*}
  |f_n(\bbeta_1) - f_n(\bbeta_2)| &= \left|\frac{w}{n} \sum_{i=1}^n \varrho\left(\frac{y_i - \bx_i^T \bbeta_1}{\hat\sigma_n}\right) - \frac{w}{n} \sum_{i=1}^n \varrho\left(\frac{y_i - \bx_i^T \bbeta_2}{\hat\sigma_n}\right)\right| \cr
  &\leq \frac{w}{n} \sum_{i=1}^n\left|\varrho\left(\frac{y_i - \bx_i^T \bbeta_1}{\hat\sigma_n}\right) - \varrho\left(\frac{y_i - \bx_i^T \bbeta_2}{\hat\sigma_n}\right)\right| \cr
  &\leq \frac{w C}{n \hat\sigma_n} \sum_{i=1}^n\left|\bx_i^T (\bbeta_1 - \bbeta_2)\right| \cr
  &\leq \frac{w C}{n \hat\sigma_n}\sum_{j=1}^p|\beta_{1j} - \beta_{2j}| \sum_{i=1}^n |x_{ij}| \cr
  &\leq \frac{w C}{\hat\sigma_n}\left(\max_j \frac{1}{n}\sum_{i=1}^n |x_{ij}|\right)\sum_{j=1}^p|\beta_{1j} - \beta_{2j}| \cr
  &\leq \frac{\sqrt{p} w C}{\sigma_0} (1 + \delta) \left(\max_j \E|X_{j}| + \delta\right)\|\bbeta_{1} - \bbeta_{2}\|.
  \end{align*}
  We used the triangle inequality in the first and third inequalities. In the second inequality, we used the fact that the function $\varrho$ is Lipschitz continuous (with, say, constant $C > 0$), because $\varrho'$ is bounded; see \autoref{sec:biweight} for the expression of $\varrho'$. In the last inequality, we used that $\sigma_0/\hat\sigma_n \rightarrow 1$ and $\frac{1}{n}\sum_{i=1}^n |x_{ij}| \rightarrow \E|X_{j}|$ as $n \rightarrow \infty$ for almost all realizations $\{\bx_i, y_i\}_{i=1}^n$ under Assumptions \ref{ass:mu_X} and \ref{ass:estimator2}. Therefore, there exists a constant $\delta > 0$ such that $\sigma_0/\hat\sigma_n \leq 1 + \delta$ and $\frac{1}{n}\sum_{i=1}^n |x_{ij}| \leq \E|X_{j}| + \delta$ for all $n$. In the last inequality, we also used the relation between the $1$-norm and the Euclidean norm.

  Now, we prove Result (b). We prove that
 \[
w \log \hat\sigma_n +  \frac{w}{n} \sum_{i=1}^n \varrho\left(\frac{Y_i - \bX_i^T \bbeta}{\hat\sigma_n}\right) \rightarrow w \log \sigma_0 + w \E\left[\varrho\left(\frac{Y - \bX^T \bbeta}{\sigma_0}\right)\right] = \bar{f}(\bbeta),
 \]
 with probability 1 as $n \rightarrow \infty$ for all $\bbeta \in \re^p$, where $\bZ_1 = (\bX_1, Y_1), \ldots, \bZ_n = (\bX_n, Y_n)$ are IID random variables such that $\bX_i \sim \mu_{\bX}$ and $Y_i = \bX_i^T \bbeta_0 + \sigma_0 \varepsilon_i$,  $\varepsilon_1, \ldots, \varepsilon_n$ being IID random variables (independent of $(\bX_1, \ldots, \bX_n)$) each having the PDF $f_0$, and $\bX \sim \mu_{\bX}$ and $Y = \bX^T \bbeta_0 + \sigma_0 \varepsilon$, $\varepsilon \sim f_0$ and independent of $\bX$. The result implies that, for almost all realizations $\{\bx_i, y_i\}_{i=1}^n$ (from the true generating model), $f_n(\bbeta) \rightarrow \bar{f}(\bbeta)$ for all $\bbeta \in \re^p$.

 We have that $w \log \hat\sigma_n \rightarrow w \log \sigma_0$ with probability 1 by \autoref{ass:estimator2}. Also,
 \begin{align*}
  &\left|\frac{w}{n} \sum_{i=1}^n \varrho\left(\frac{Y_i - \bX_i^T \bbeta}{\hat\sigma_n}\right) - w \E\left[\varrho\left(\frac{Y - \bX^T \bbeta}{\sigma_0}\right)\right]\right| \cr
   & \quad\leq \left|\frac{w}{n} \sum_{i=1}^n \varrho\left(\frac{Y_i - \bX_i^T \bbeta}{\hat\sigma_n}\right) - \frac{w}{n} \sum_{i=1}^n \varrho\left(\frac{Y_i - \bX_i^T \bbeta}{\sigma_0}\right)\right| \cr
   &\qquad+ \left|\frac{w}{n} \sum_{i=1}^n \varrho\left(\frac{Y_i - \bX_i^T \bbeta}{\sigma_0}\right) - w \E\left[\varrho\left(\frac{Y - \bX^T \bbeta}{\sigma_0}\right)\right]\right|,
 \end{align*}
 by the triangle inequality.

 We have that $0 \leq \varrho \leq 1$, implying that
 \[
  0 \leq  \E\left[\varrho\left(\frac{Y - \bX^T \bbeta}{\sigma_0}\right)\right] \leq 1,
 \]
 which in turn implies that
 \[
 \left|\frac{w}{n} \sum_{i=1}^n\varrho\left(\frac{Y_i - \bX_i^T \bbeta}{\sigma_0}\right) -  w \E\left[\varrho\left(\frac{Y - \bX^T \bbeta}{\sigma_0}\right)\right]\right| \rightarrow 0
\]
with probability 1 by the strong law of large numbers.

We have that,
\begin{align*}
 &\left|\frac{w}{n} \sum_{i=1}^n \left(\varrho\left(\frac{Y_i - \bX_i^T \bbeta}{\hat\sigma_n}\right) - \left(\varrho\left(\frac{Y_i - \bX_i^T \bbeta}{\sigma_0}\right)\right)\right)\right| \cr
 &\quad \leq \frac{w}{n} \sum_{i=1}^n \left| \varrho\left(\frac{Y_i - \bX_i^T \bbeta}{\hat\sigma_n}\right) -\varrho\left(\frac{Y_i - \bX_i^T \bbeta}{\sigma_0}\right)\right| \cr
 &\quad \leq C \, \frac{w}{n} \sum_{i=1}^n \left|\frac{Y_i - \bX_i^T \bbeta}{\hat\sigma_n} - \frac{Y_i - \bX_i^T \bbeta}{\sigma_0}\right| \cr
 &\quad \leq C \, \frac{|\hat\sigma_n - \sigma_0|}{\hat\sigma_n} \frac{w}{n} \sum_{i=1}^n \left|\frac{Y_i - \bX_i^T \bbeta}{\sigma_0}\right|,
\end{align*}
using the triangle inequality in the first inequality and, in the second equality, the fact that function $\varrho$ is Lipschitz continuous, as before. We have that $\hat\sigma_n > 0$ and $|\hat\sigma_n - \sigma_0| \rightarrow 0$ with probability 1 under \autoref{ass:estimator2}. Therefore, the last term above goes to 0 if, for all $\bbeta \in \re^p$,
\[
 \E\left|\frac{Y - \bX^T \bbeta}{\sigma_0}\right| < \infty
\]
because, in this case,
\[
 \frac{1}{n} \sum_{i=1}^n \left|\frac{Y_i - \bX_i^T \bbeta}{\sigma_0}\right| \rightarrow \E\left|\frac{Y - \bX^T \bbeta}{\sigma_0}\right|
\]
with probability 1 by the strong law of large numbers. Using the triangle inequality twice,
\begin{align*}
 \E\left|\frac{Y - \bX^T \bbeta}{\sigma_0}\right| &= \E\left|\frac{Y  - \bX^T \bbeta_0 + \bX^T \bbeta_0 - \bX^T \bbeta}{\sigma_0}\right| \cr
 &\leq\E|\varepsilon| + \frac{1}{\sigma_0} \, \E|\bX^T(\bbeta_0 - \bbeta)| \cr
 &\leq \E|\varepsilon| + \frac{1}{\sigma_0} \sum_{j=1}^p |\beta_{0j} - \beta_j| \, \E|X_j|.
\end{align*}
Under Assumptions \ref{ass:error} and \ref{ass:mu_X}, $\E|\varepsilon| < \infty$ and $\E|X_j| < \infty$ for all $j$. This concludes the proof that $f_n(\bbeta) \rightarrow \bar{f}(\bbeta)$ for all $\bbeta \in \re^p$ and for almost all realizations $\{\bx_i, y_i\}_{i=1}^n$. Note that it is possible to prove that $f_n(\hat\bbeta_n) \rightarrow \bar{f}(\bbeta_0)$, for almost all realizations $\{\bx_i, y_i\}_{i=1}^n$, by combining the strategy above with ideas from the proof of
\[
 \frac{w}{\hat\sigma_n^2} \frac{1}{n} \sum_{i=1}^n \varrho''\left(\frac{Y_i - \bX_i^T \hat\bbeta_n}{\hat\sigma_n}\right) \bX_i \bX_i^T \rightarrow \bH_0
\]
that is presented at the beginning of the proof of \autoref{Thm:large-sample}.

We now prove Result (c). We have that
\begin{align*}
 \bar{f}(\bbeta) &= w \log \sigma_0 + w \E\left[\varrho\left(\frac{Y - \bX^T \bbeta}{\sigma_0}\right)\right] \cr
 &= w \log \sigma_0 + w \E\left[\varrho\left(\frac{Y - \bX^T \bbeta_0}{\sigma_0} - \bX^T\left(\frac{\bbeta - \bbeta_0}{\sigma_0}\right)\right)\right] \cr
 &= w \log \sigma_0 +  w \E\left[\varrho\left(\varepsilon - \bX^T \bD\right)\right],
\end{align*}
using the reparametrization $\bD = (\bbeta - \bbeta_0) / \sigma_0$. In the following, we view the function $\bar{f}$ as a function of $\bD \in \re^p$.

We have that
\begin{align*}
 \E\left[\varrho\left(\varepsilon - \bX^T \bD\right)\right] = \int \mu_{\bX}(\d\bx)\int\varrho\left(\varepsilon - \bx^T \bD\right)  f_0(\varepsilon) \, \d\varepsilon, \quad \bD \in \re^p.
\end{align*}
Let us consider that $\bD \neq \mathbf{0}$. Because the components in $\bX$ (except the first one) are continuous random variables under \autoref{ass:mu_X}, we know that the event $\bX^T \bD = 0$ has probability 0. Therefore, for $\mu_{\bX}$-almost all $\bx$,
\[
    \int \varrho\left(\varepsilon - \bx^T \bD\right)  f_0(\varepsilon) \, \d\varepsilon > \int \varrho\left(\varepsilon\right)  f_0(\varepsilon) \, \d\varepsilon,
\]
when $\bD \neq \mathbf{0}$ under \autoref{ass:error} by \autoref{lemma1}. This proves that
\[
 \bar{f}(\bD) = w \log \sigma_0 +  w \E\left[\varrho\left(\varepsilon - \bX^T \bD\right)\right] > w \log \sigma_0 +  w \E\left[\varrho\left(\varepsilon\right)\right] = \bar{f}(\mathbf{0}),
\]
for all $\bD \neq \mathbf{0}$ and that $\bar{f}$ is minimized at $\bD = \mathbf{0}$ (or equivalently $\bbeta = \bbeta_0$ in the original parametrization). This minimum is unique and global.

Finally, we prove Result (d). Under \autoref{ass:estimator2}, we have that
\[
 \liminf_n \inf_{\bbeta \in B_C(\hat\bbeta_n)^\mathsf{c}} (f_n(\bbeta) - f_n(\hat\bbeta_n)) > 0
\]
 for almost all realizations $\{\bx_i, y_i\}_{i=1}^n$ (from the true generating model) and a constant $C > 0$. There thus remains to prove that $\liminf_n \inf_{\bbeta: r \leq \|\bbeta - \hat\bbeta_n\| \leq C} (f_n(\bbeta) - f_n(\hat\bbeta_n)) > 0$  for almost all realizations $\{\bx_i, y_i\}_{i=1}^n$ and $r \in (0, C)$. The proof holds for almost all realizations $\{\bx_i, y_i\}_{i=1}^n$ and, therefore, we omit repeating this for simplicity.

 As a consequence of Results (a) and (b), we have that $f_n$ converges uniformly to $\bar{f}$ as $n \rightarrow \infty$ on any compact subset of $\re^p$. Using that $\hat\bbeta_n \rightarrow \bbeta_0$ as $n \rightarrow \infty$ with probability 1 under \autoref{ass:estimator2}, we take a compact subset $A \subset \re^p$ such that $\{\bbeta: r \leq \|\bbeta - \hat\bbeta_n\| \leq C\}$ is included in it for all large enough $n$ and such that $\|\bbeta - \bbeta_0\| \geq r$ for all $\bbeta \in A$.

 As a consequence of Result (c), we have that $\bar{f}(\bbeta) - \bar{f}(\bbeta_0) > 0$ for all $\bbeta \in A$. In fact, given the definition of $A$, we can find a small enough $\delta > 0$ such that $\inf_{\bbeta \in A}(\bar{f}(\bbeta) - \bar{f}(\bbeta_0) - 2\delta) > 0$. For this $\delta$, we have $f_n(\bbeta) - f_n(\hat\bbeta_n) > \bar{f}(\bbeta) - \bar{f}(\bbeta_0) - 2\delta > 0$ for large enough $n$, uniformly on $A$. Therefore, $\liminf_n \inf_{\bbeta: r \leq \|\bbeta - \hat\bbeta_n\| \leq C} (f_n(\bbeta) - f_n(\hat\bbeta_n)) > 0$.
 \end{proof}

 \begin{proof}[Proof of \autoref{Thm:large-sample}]
    The results follow from the application of Theorem 4 in \cite{miller2021asymptotic}. Under the assumptions of \autoref{Thm:large-sample} and using \autoref{lemma:large-sample}, we need to prove the following to apply Theorem 4 in \cite{miller2021asymptotic}:
 \begin{align}\label{eqn:Taylor}
  f_n(\bbeta) = f_n(\hat{\bbeta}_n) + \frac{1}{2}(\bbeta - \hat\bbeta_n)^T \bH_n (\bbeta - \hat\bbeta_n) + r_n(\bbeta - \hat\bbeta_n), \quad \bbeta \in \re^p,
 \end{align}
 where $\bH_n$ is a symmetric $p \times p$ matrix such that $\bH_n \rightarrow \bH_0$ for almost all realizations $\{\bx_i, y_i\}_{i=1}^n$ (from the true generating model) and $r_n: \re^p \rightarrow \re$ has the following property: there exist $c, r > 0$ such that for all $n$ sufficiently large and for all $\bx \in B_r(\mathbf{0})$, $|r_n(\bx)| \leq c \|\bx\|^3$. Note that the convergence of a matrix is with respect to the Frobenius norm.

 To prove the above, we apply a Taylor expansion on $f_n$ around $\hat\bbeta_n$:
  \begin{align*}
  f_n(\bbeta) = f_n(\hat\bbeta_n)  &+ f_n'(\hat\bbeta_n)^T (\bbeta - \hat\bbeta_n)+ \frac{1}{2}(\bbeta - \hat\bbeta_n)^T f_n''( \hat\bbeta_n) (\bbeta - \hat\bbeta_n) \cr
  &\quad+ \frac{1}{6}\sum_{j = 1}^p \sum_{k = 1}^p \sum_{s = 1}^p \frac{\partial^3 f_n(\tilde\bbeta)}{\partial \beta_j \partial \beta_k \partial \beta_s} (\beta_j - \tilde{\beta}_j)(\beta_k - \tilde{\beta}_k)(\beta_s - \tilde{\beta}_s),
 \end{align*}
 where $f_n'$ and $f_n''$ are the gradient and Hessian of $f_n$, respectively, and $\tilde{\beta}$ is on the line segment connecting $\bbeta$ and $\hat\bbeta_n$.

 First, $f_n'(\bbeta_n) = \mathbf{0}$ for all $n$ under \autoref{ass:estimator2}. Second, we have that
 \[
  f_n''(\hat\bbeta_n) = \frac{w}{\hat\sigma_n^2} \frac{1}{n} \sum_{i=1}^n \varrho''\left(\frac{y_i - \bx_i^T \hat\bbeta_n}{\hat\sigma_n}\right) \bx_i \bx_i^T,
 \]
 where $\varrho''$ is the second derivative of $\varrho = -\log g$ (see \autoref{sec:biweight} for the expression of $\varrho''$). The matrix $f_n''( \bbeta_n)$ is $p\times p$ and symmetric. We now prove that
 \begin{align*}
  \frac{w}{\hat\sigma_n^2} \frac{1}{n} \sum_{i=1}^n \varrho''\left(\frac{Y_i - \bX_i^T \hat\bbeta_n}{\hat\sigma_n}\right) \bX_i \bX_i^T &\rightarrow \frac{w}{\sigma_0^2} \E\left[\varrho''\left(\frac{Y - \bX^T \bbeta_0}{\sigma_0}\right) \bX \bX^T\right] \cr
   &= \frac{w}{\sigma_0^2} \E[\varrho''(\varepsilon)] \E[\bX \bX^T] = \bH_0,
 \end{align*}
  with probability 1 as $n \rightarrow \infty$, where $\bZ_1 = (\bX_1, Y_1), \ldots, \bZ_n = (\bX_n, Y_n)$ are IID random variables such that $\bX_i \sim \mu_{\bX}$ and $Y_i = \bX_i^T \bbeta_0 + \sigma_0 \varepsilon_i$,  $\varepsilon_1, \ldots, \varepsilon_n$ being IID random variables (independent of $(\bX_1, \ldots, \bX_n)$) each having the PDF $f_0$, and $\bX \sim \mu_{\bX}$ and $Y = \bX^T \bbeta_0 + \sigma_0 \varepsilon$, $\varepsilon \sim f_0$ and independent of $\bX$. This implies that, for almost all realizations $\{\bx_i, y_i\}_{i=1}^n$ (from the true generating model),
  \[
  \bH_n = f_n''(\hat\bbeta_n) = \frac{w}{\hat\sigma_n^2} \frac{1}{n} \sum_{i=1}^n \varrho''\left(\frac{y_i - \bx_i^T \hat\bbeta_n}{\hat\sigma_n}\right) \bx_i \bx_i^T \rightarrow \frac{w}{\sigma_0^2} \E[\varrho''(\varepsilon)] \E[\bX \bX^T] = \bH_0.
 \]
 Note that Lemmas \ref{lemma:varrhopprim} and \ref{lemma:positive} ensure that $\bH_0$ is a positive definite matrix.

By \autoref{ass:estimator2}, $w / \hat\sigma_n^2 \rightarrow w / \sigma_0^2$ with probability 1. To prove that
\[
 \frac{1}{n} \sum_{i=1}^n \varrho''\left(\frac{Y_i - \bX_i^T \hat\bbeta_n}{\hat\sigma_n}\right) \bX_i \bX_i^T \rightarrow \E[\varrho''(\varepsilon)] \E[\bX \bX^T],
\]
we show the convergence element by element. Let us analyse the element at the $j$-th line and $k$-th column:
\begin{align*}
 &\left|\frac{1}{n} \sum_{i=1}^n \varrho''\left(\frac{Y_i - \bX_i^T \hat\bbeta_n}{\hat\sigma_n}\right) X_{ij} X_{ik} - \E[\varrho''(\varepsilon)] \E[X_j X_k]\right| \cr
 &\quad\leq \left|\frac{1}{n} \sum_{i=1}^n \varrho''\left(\frac{Y_i - \bX_i^T \hat\bbeta_n}{\hat\sigma_n}\right) X_{ij} X_{ik} - \frac{1}{n} \sum_{i=1}^n \varrho''\left(\frac{Y_i - \bX_i^T \bbeta_0}{\sigma_0}\right) X_{ij} X_{ik}\right| \cr
 &\qquad + \left|\frac{1}{n} \sum_{i=1}^n \varrho''\left(\frac{Y_i - \bX_i^T \bbeta_0}{\sigma_0}\right) X_{ij} X_{ik} - \E[\varrho''(\varepsilon)] \E[X_j X_k]\right|,
\end{align*}
using the triangle inequality. As the expectations $\E[\varrho''(\varepsilon)]$ and $\E[X_j X_k]$ exist ($|\varrho''|$ is bounded and $\E[X_j X_k]$ exists by \autoref{ass:mu_X} because $\E|X_j X_k|\leq \E[X_j^2]^{1/2}\E[X_k^2]^{1/2}$ using the Cauchy--Schwarz inequality),
\[
 \left|\frac{1}{n} \sum_{i=1}^n \varrho''\left(\frac{Y_i - \bX_i^T \bbeta_0}{\sigma_0}\right) X_{ij} X_{ik} - \E[\varrho''(\varepsilon)] \E[X_j X_k]\right| \rightarrow 0,
\]
with probability 1 as $n \rightarrow \infty$ by the strong law of large numbers (using that $(Y - \bX^T\beta_0) / \sigma_0 = \varepsilon$ and the independence between $\varepsilon$ and $\bX$).

We analyse the other term:
\begin{align*}
 &\left|\frac{1}{n} \sum_{i=1}^n \varrho''\left(\frac{Y_i - \bX_i^T \hat\bbeta_n}{\hat\sigma_n}\right) X_{ij} X_{ik} - \frac{1}{n} \sum_{i=1}^n \varrho''\left(\frac{Y_i - \bX_i^T \bbeta_0}{\sigma_0}\right) X_{ij} X_{ik}\right| \cr
 &\quad \leq \frac{1}{n} \sum_{i=1}^n|X_{ij} X_{ik}|\left|\varrho''\left(\frac{Y_i - \bX_i^T \hat\bbeta_n}{\hat\sigma_n}\right) - \varrho''\left(\frac{Y_i - \bX_i^T \bbeta_0}{\sigma_0}\right)\right| \cr
 &\quad \leq \frac{C}{n} \sum_{i=1}^n|X_{ij} X_{ik}|\left|\frac{Y_i - \bX_i^T \hat\bbeta_n}{\hat\sigma_n} - \frac{Y_i - \bX_i^T \bbeta_0}{\sigma_0}\right| \cr
 &\quad \leq  \frac{C}{n} \sum_{i=1}^n|X_{ij} X_{ik}|\left|\frac{Y_i}{\hat\sigma_n} - \frac{Y_i }{\sigma_0}\right| + \frac{C}{n} \sum_{i=1}^n|X_{ij} X_{ik}|\left|\bX_i^T\left(\frac{\hat\bbeta_n}{\hat\sigma_n} - \frac{\bbeta_0 }{\sigma_0}\right)\right|  \cr
  &\quad \leq  \left|\frac{1}{\hat\sigma_n} - \frac{1}{\sigma_0}\right|\frac{C}{n} \sum_{i=1}^n|X_{ij} X_{ik}||Y_i| + \sum_{s=1}^p \left|\frac{\hat\beta_{ns}}{\hat\sigma_n} - \frac{\beta_{0s}}{\sigma_0}\right| \frac{C}{n} \sum_{i=1}^n|X_{ij} X_{ik}X_{is}|,
\end{align*}
using the triangle inequality in the first inequality, the fact that $\varrho''$ is Lipschitz continuous in the second inequality (say, with
constant $C > 0$) because its derivative is bounded (see \autoref{sec:biweight} for the expression of $\varrho'''$), and again the triangle inequality in the last two inequalities.

As $n \rightarrow \infty$, $|1/\hat\sigma_n - 1/\sigma_0| \rightarrow 0$ and $|\hat\beta_{ns}/ \hat\sigma_n - \beta_{0s} / \sigma_0|\rightarrow 0$ for all $s$, with probability 1, by \autoref{ass:estimator2}. By the strong law of large numbers, we thus have that the sum above converges to 0 with probability 1 as $n \rightarrow \infty$ if $\E|X_{j} X_{k}||Y| < \infty$ and $\E|X_{j} X_{k}X_{s}| < \infty$. Using the triangle inequality twice,
\begin{align*}
 \E|X_{j} X_{k}||Y| = \E|X_{j} X_{k}||\bX^T\bbeta_0 + \sigma_0 \varepsilon| \leq \sum_{s=1}^p |\beta_{0s}| \E|X_jX_kX_s| + \sigma_0\E|\varepsilon|\E|X_{j} X_{k}|.
\end{align*}
Under \autoref{ass:error}, $\E|\varepsilon| < \infty$, and we know that $\E|X_{j} X_{k}| < \infty$ under \autoref{ass:mu_X}. There thus remains to prove that $\E|X_{j} X_{k}X_{s}|<\infty$. Applying the generalized Hölder's inequality yields
\begin{align*}
 \E|X_{j} X_{k}X_{s}|\leq [\E|X_{j}|^3]^{1/3} [\E|X_{k}|^3]^{1/3} [\E|X_{s}|^3]^{1/3},
\end{align*}
which is finite under \autoref{ass:mu_X}.

To prove \eqref{eqn:Taylor}, there thus remains to show that
\[
 \left|\sum_{j = 1}^p \sum_{k = 1}^p \sum_{s = 1}^p \frac{\partial^3 f_n(\tilde\bbeta)}{\partial \beta_j \partial \beta_k \partial \beta_s} (\beta_j - \tilde{\beta}_j)(\beta_k - \tilde{\beta}_k)(\beta_s - \tilde{\beta}_s)\right| \leq c \|\bbeta - \tilde{\bbeta}\|^3,
\]
for a constant $c > 0$, recalling that $\tilde{\bbeta}$ is on the line segment connecting $\bbeta$ and $\hat\bbeta_n$. We have that
 \begin{align*}
  \left|\frac{\partial^3 f_n(\tilde\bbeta)}{\partial \beta_j \partial \beta_k \partial \beta_s}\right| &= \left|\frac{w}{\hat\sigma_n^3} \frac{1}{n} \sum_{i=1}^n \varrho'''\left(\frac{y_i - \bx_i^T \tilde\bbeta}{\hat\sigma_n}\right) x_{ij}x_{ik}x_{is}\right| \cr
  &\leq \frac{w}{\hat\sigma_n^3} \frac{1}{n} \sum_{i=1}^n \left|\varrho'''\left(\frac{y_i - \bx_i^T \tilde\bbeta}{\hat\sigma_n}\right) x_{ij}x_{ik}x_{is}\right| \cr
  &\leq \frac{w C'}{\hat\sigma_n^3} \frac{1}{n} \sum_{i=1}^n |x_{ij}x_{ik}x_{is}| \cr
  &\leq \frac{w C'}{\sigma_0^3}(1+\delta)^3(\max_{j,k,s}\E|X_{j}X_{k}X_{s}| + \delta).
 \end{align*}
 In the first inequality, we used the triangle inequality. In the second inequality, we used that $|\varrho'''|$ is bounded (by, say, a constant $C' > 0$; see \autoref{sec:biweight} for the expression of $\varrho'''$). In the last inequality, we used that $\sigma_0 / \hat\sigma_n \rightarrow 1$ and $\frac{1}{n} \sum_{i=1}^n |x_{ij}x_{ik}x_{is}| \rightarrow \E|X_{j}X_{k}X_{s}|$ as $n \rightarrow \infty$ for almost all realizations $\{\bx_i, y_i\}_{i=1}^n$ under Assumptions \ref{ass:mu_X} and \ref{ass:estimator2}. Therefore, there exists a constant $\delta > 0$ such that $\sigma_0 / \hat\sigma_n \leq 1 + \delta$ and $\frac{1}{n} \sum_{i=1}^n |x_{ij}x_{ik}x_{is}| \leq \E|X_{j}X_{k}X_{s}| + \delta$ for all $n$ (recall that $\E|X_{j}X_{k}X_{s}| < \infty$ for all $j, k$ and $s$ as proved above). Consequently, the third derivative of $f_n$ is bounded, by, say, a constant $C > 0$, for almost all realizations $\{\bx_i, y_i\}_{i=1}^n$ (from the true generating model), and
 \begin{align*}
& \left|\sum_{j = 1}^p \sum_{k = 1}^p \sum_{s = 1}^p \frac{\partial^3 f_n(\tilde\bbeta)}{\partial \beta_j \partial \beta_k \partial \beta_s} (\beta_j - \tilde{\beta}_j)(\beta_k - \tilde{\beta}_k)(\beta_s - \tilde{\beta}_s)\right| \cr
&\quad \leq C \|\bbeta - \tilde\bbeta\|_1^3 \leq Cp^{3/2} \|\bbeta - \tilde\bbeta\|^3,
 \end{align*}
using the triangle inequality in the first inequality and the relation between the $1$-norm and the Euclidean norm in the second inequality ($\| \, \cdot \,\|_1$ denoting the $1$-norm). This concludes the proof.
\end{proof}

\begin{Lemma}[Global minimum of a convolution]\label{lemma1}
    Let $h, f: \re \rightarrow \re$ be two functions such that both are symmetric with respect to 0, $\tilde{h} := h(| \, \cdot \,|): [0, \infty) \rightarrow \re$ is strictly increasing up to $k > 0$ and $\tilde{f} := f(| \, \cdot \,|): [0, \infty) \rightarrow \re$ is strictly decreasing, that is for any $|y| \geq |x|$, $f(y) = f(|y|) < f(|x|) = f(x)$ and $h(y) = h(|y|) > h(|x|) = h(x)$ if $|x| < k$ and $h(y) \geq h(x)$ if $|x| \geq k$. Then, the convolution $(h * f):\re \rightarrow \re$ defined as
    \[
     (h * f)(y) = \int_{-\infty}^\infty h(x - y) f(x) \, \d x
    \]
    has a unique global minimum at $0$. The result also holds if $\tilde{h}$ is strictly increasing, that is $h(y) > h(x)$ for any $|y| \geq |x|$.
\end{Lemma}

\begin{proof}
Let $y \neq 0$. We prove that
\[
 \int_{-\infty}^\infty (h(x - y) - h(x)) f(x) \, \d x > 0.
\]
Let us consider that $y < 0$. The proof for the case $y > 0$ is analogous. We have that
\begin{align*}
 &\int_{-\infty}^\infty (h(x - y) - h(x)) f(x) \, \d x \cr
  &\quad= \int_{y}^\infty (h(x - y) - h(x)) f(x) \, \d x - \int_{-\infty}^{y} (h(x) - h(x - y)) f(x) \, \d x.
\end{align*}

We obtain an upper bound for the second integral on the right-hand side (RHS):
\begin{align*}
 \int_{-\infty}^{y} (h(x) - h(x - y)) f(x) \, \d x &= \int_{-y}^{\infty} (h(u) - h(-u - y)) f(u) \, \d u \cr
  &< \int_{-y}^{\infty} (h(u) - h(u + y)) f(u + y) \, \d u \cr
 & = \int_{0}^{\infty} (h(z - y) - h(z)) f(z) \, \d z.
\end{align*}
In the first equality, we used the change of variable $u = -x$. In the inequality, we used that $u > u + y \geq 0$, implying that $h(u) - h(u + y) > 0$ when $0 \leq u + y < k$ and $h(u) - h(u + y) \geq 0$ when $u + y \geq k$, and $f(u) < f(u + y)$. In the last equality, we used the change of variable $z = u + y \Leftrightarrow z - y= u$.

Therefore,
\begin{align*}
 &\int_{-\infty}^\infty (h(x - y) - h(x)) f(x) \, \d x \cr
 &\quad> \int_{y}^\infty (h(x - y) - h(x)) f(x) \, \d x - \int_{0}^{\infty} (h(z - y) - h(z)) f(z) \, \d z \cr
 &\quad = \int_{y}^0 (h(x - y) - h(x)) f(x) \, \d x.
\end{align*}

We proceed similarly as before to show that this integral is greater than or equal to $0$:
\begin{align*}
 \int_{y}^0 (h(x - y) - h(x)) f(x) \, \d x &= \int_{y / 2}^{-y / 2} (h(u - y / 2) - h(u + y / 2)) f(u + y / 2) \, \d u \cr
 & = \int_{0}^{-y / 2} (h(u - y / 2) - h(u + y / 2)) f(u + y / 2) \, \d u \cr
 &\quad - \int_{y / 2}^{0} (h(u + y / 2) - h(u - y / 2)) f(u + y / 2) \, \d u,
\end{align*}
using the change of variable $u = x - y / 2 \Leftrightarrow x = u + y/2$. We obtain an upper bound for the second integral on the RHS:
\begin{align*}
 &\int_{y / 2}^{0} (h(u + y / 2) - h(u - y / 2)) f(u + y / 2) \, \d u \cr
  &\quad= \int_{0}^{-y / 2} (h(z - y / 2) - h(z + y / 2)) f(z - y / 2) \, \d z \cr
&\quad \leq \int_{0}^{-y / 2} (h(z - y / 2) - h(z + y / 2)) f(z + y / 2) \, \d z.
\end{align*}
In the first equality, we used the change of variable $z = -u$. In the inequality, we used that $|z - y / 2| \geq |z + y / 2|$ when $-y / 2 \geq z \geq 0$, implying that $h(z - y / 2) - h(z + y / 2) \geq 0$ and $f(z - y / 2) < f(z + y / 2)$. Indeed,
\[
 (z - y / 2)^2 = (z + y / 2 - y)^2 = (z + y / 2)^2 - 2 y z \geq (z + y / 2)^2,
\]
because $y^2 \geq - 2 y z \geq 0$. This concludes the proof. Note that the same proof can be used if $\tilde{h}$ is strictly increasing.
\end{proof}

\begin{Lemma}\label{lemma:varrhopprim}
Let $\varepsilon \sim f_0$. If $f_0$ is symmetric with respect to the origin, strictly decreasing from the origin and continuously differentiable, then $\E[\varrho''(\varepsilon)] \in (0, \infty)$, where $\varrho''$ is the second derivative of $\varrho$ (see \autoref{sec:biweight} for the expression of $\varrho''$).
\end{Lemma}

\begin{proof}
    First, $\E[|\varrho''(\varepsilon)|] < \infty$ given that $\varrho''$ is bounded, implying that $\E[\varrho''(\varepsilon)] < \infty$. Now, we prove that $\E[\varrho''(\varepsilon)] > 0$. The function $\varrho''$ is symmetric with respect to the origin and it is null for all $|\varepsilon| > k$. Using that $f_0$ is also symmetric with respect to the origin,
 \[
  \E[\varrho''(\varepsilon)] = 2 \int_0^k \varrho''(\varepsilon) \, f_0(\varepsilon) \, \d\varepsilon.
 \]

 Integration by parts yields
 \[
  \int_0^k \varrho''(\varepsilon) \, f_0(\varepsilon) \, \d\varepsilon = \int_0^k \varrho'(\varepsilon) \, (-f_0'(\varepsilon)) \, \d\varepsilon > 0,
 \]
 given that $\varrho'(0) = \varrho'(k) = 0$ (see \autoref{sec:biweight} for the expression of $\varrho'$) and that $\varrho'(\varepsilon) \, (-f_0'(\varepsilon)) > 0$ for all $\varepsilon \in (0, k)$. This concludes the proof.
\end{proof}

It is relatively well known that $\E[\bX \bX^T]$ is positive semi-definite in general. Below we present a lemma stating that in the case where the components in $\bX \sim \mu_{\bX}$ (except the first one) are continuous random variables, $\E[\bX \bX^T]$ is positive definite.

\begin{Lemma}\label{lemma:positive}
 Assume that the components in $\bX \sim \mu_{\bX}$ (except the first one) are continuous random variables. Then, $\E[\bX \bX^T]$ is positive definite.
\end{Lemma}

\begin{proof}[Proof of \autoref{lemma:positive}]
 Let $\boldsymbol\omega \in \re^p$ with $\boldsymbol\omega \neq \mathbf{0}$. We have that
 \begin{align*}
  \boldsymbol\omega^T \E[\bX \bX^T] \boldsymbol\omega = \E[\boldsymbol\omega^T \bX \bX^T \boldsymbol\omega] = \E[(\boldsymbol\omega^T \bX)^2].
 \end{align*}
 Given that the components in $\bX \sim \mu_{\bX}$ (except the first one) are continuous random variables (by assumption) and $\boldsymbol\omega \neq \mathbf{0}$, the event $\boldsymbol\omega^T \bX = 0$ has probability 0. Therefore, $(\boldsymbol\omega^T \bX)^2 > 0$ with probability 1, implying that $\E[(\boldsymbol\omega^T \bX)^2] = \boldsymbol\omega^T \E[\bX \bX^T] \boldsymbol\omega > 0$.
\end{proof}

\section{About estimation of Bayesian models}\label{sec:info_estimation}

\subsection{Normal model}

In Sections \ref{sec:example1} and \ref{sec:example_reserve}, we present estimation results for the Bayesian normal linear regression based on the data sets without the outliers. The following proposition allows to obtain such results. It is relatively well known. We provide the proof for completeness.

\begin{Proposition}\label{prop:post_norm}
   In the model in \eqref{eqn:linear_reg}, if $f$ is the standard normal PDF and $\pi$ is such that $\bbeta$ given $\xi = \sigma^2$ has a normal distribution with a mean of $\bmu_{\bbeta} \in \re^p$ and a (positive definite) covariance matrix of $\xi \bSigma_{\bbeta}$ and $\xi$ has an inverse-gamma distribution with a shape parameter of $a > 0$ and a scale parameter of $b > 0$, then the posterior distribution is such that: $\bbeta$ given $\xi$ has a normal distribution with a mean of $\hat{\bbeta}$ and a covariance matrix of $\xi(\bX^T \bX + \bLambda_{\bbeta})^{-1}$, and $\xi$ has an inverse-gamma distribution with a shape parameter of $(2a + n) / 2$ and a scale parameter of
   \[
 \frac{2b + \by^T \by  - \hat{\bbeta}^T (\bX^T \bX + \bLambda_{\bbeta}) \hat{\bbeta} + \bmu_{\bbeta}^T \bLambda_{\bbeta} \bmu_{\bbeta}}{2},
\]
 where $\hat{\bbeta} = (\bX^T \bX + \bLambda_{\bbeta})^{-1} (\bX^T \by + \bLambda_{\bbeta} \bmu_{\bbeta})$ and $\bLambda_{\bbeta} = \bSigma_{\bbeta}^{-1}$.
\end{Proposition}

\begin{proof}
  In normal linear regression, $\bY$, given $\bbeta$ and $\xi$, has a normal distribution with a mean of $\bX \bbeta$ and a covariance matrix of $\xi \I_n$, where $\I_n$ is the identity matrix of size $n$. Therefore, we can write the posterior density as:
 \begin{align*}
  &\pi(\bbeta, \xi \mid \by) \cr
  &\propto  \pi(\xi) \, \frac{1}{\xi^{p/2}} \exp\left(-\frac{1}{2\xi} \, (\bbeta - \bmu_{\bbeta})^T \bLambda_{\bbeta} (\bbeta - \bmu_{\bbeta})\right)  \frac{1}{\xi^{n/2}} \exp\left(-\frac{1}{2\xi} (\by - \bX \bbeta)^T (\by - \bX \bbeta)\right) \cr
  &= \pi(\xi) \, \frac{1}{\xi^{(p + n) / 2}} \exp\left(-\frac{1}{2\xi} \left[(\bbeta - \bmu_{\bbeta})^T \bLambda_{\bbeta} (\bbeta - \bmu_{\bbeta}) + (\by - \bX \bbeta)^T (\by - \bX \bbeta)\right] \right).
 \end{align*}
 We analyse the term in the exponential:
 \begin{align*}
 & (\by - \bX \bbeta)^T (\by - \bX \bbeta) + (\bbeta - \bmu_{\bbeta})^T \bLambda_{\bbeta} (\bbeta - \bmu_{\bbeta}) \cr
 &\quad= \by^T \by - \by^T \bX \bbeta - (\bX \bbeta)^T \by + \bbeta^T \bX^T \bX \bbeta + (\bbeta - \bmu_{\bbeta})^T \bLambda_{\bbeta} (\bbeta - \bmu_{\bbeta})\cr
  &\quad= \by^T \by - 2\by^T \bX \bbeta + (\bbeta - \hat{\bbeta} + \hat{\bbeta})^T \bX^T \bX (\bbeta - \hat{\bbeta} + \hat{\bbeta}) \cr
  &\qquad + (\bbeta - \hat{\bbeta} + \hat{\bbeta}  - \bmu_{\bbeta})^T \bLambda_{\bbeta} (\bbeta - \hat{\bbeta} + \hat{\bbeta} - \bmu_{\bbeta})\cr
  &\quad= \by^T \by - 2\by^T \bX \bbeta + (\bbeta - \hat{\bbeta})^T \bX^T \bX (\bbeta - \hat{\bbeta}) \cr
  &\qquad +  (\bbeta - \hat{\bbeta})^T \bX^T \bX \hat{\bbeta} +  \hat{\bbeta}^T \bX^T \bX \bbeta \cr
  &\qquad + (\bbeta - \hat{\bbeta})^T \bLambda_{\bbeta} (\bbeta - \hat{\bbeta})\cr
  &\qquad +(\bbeta - \hat{\bbeta})^T \bLambda_{\bbeta} (\hat{\bbeta} - \bmu_{\bbeta}) + (\hat{\bbeta} - \bmu_{\bbeta})^T \bLambda_{\bbeta} (\bbeta - \bmu_{\bbeta})\cr
  &\quad= \by^T \by - 2\by^T \bX \bbeta + (\bbeta - \hat{\bbeta})^T (\bX^T \bX + \bLambda_{\bbeta}) (\bbeta - \hat{\bbeta}) \cr
    &\qquad +2 \bbeta^T \bX^T \bX \hat{\bbeta} - \hat{\bbeta}^T \bX^T \bX \hat{\bbeta}  \cr
  &\qquad +(\bbeta - \hat{\bbeta})^T \bLambda_{\bbeta} (\hat{\bbeta} - \bmu_{\bbeta}) + (\hat{\bbeta} - \bmu_{\bbeta})^T \bLambda_{\bbeta} (\bbeta - \bmu_{\bbeta})\cr
  &\quad= \by^T \by - 2\by^T \bX \bbeta + (\bbeta - \hat{\bbeta})^T (\bX^T \bX + \bLambda_{\bbeta}) (\bbeta - \hat{\bbeta}) \cr
    &\qquad +2 \bbeta^T (\bX^T \bX + \bLambda_{\bbeta} - \bLambda_{\bbeta})  \hat{\bbeta} - \hat{\bbeta}^T (\bX^T \bX + \bLambda_{\bbeta} - \bLambda_{\bbeta}) \hat{\bbeta}  \cr
  &\qquad +(\bbeta - \hat{\bbeta})^T \bLambda_{\bbeta} (\hat{\bbeta} - \bmu_{\bbeta}) + (\hat{\bbeta} - \bmu_{\bbeta})^T \bLambda_{\bbeta} (\bbeta - \bmu_{\bbeta})\cr
  &\quad= \by^T \by + (\bbeta - \hat{\bbeta})^T (\bX^T \bX + \bLambda_{\bbeta}) (\bbeta - \hat{\bbeta}) - \hat{\bbeta}^T(\bX^T \bX + \bLambda_{\bbeta}) \hat{\bbeta} \cr
    &\qquad +2 \bbeta^T \bLambda_{\bbeta} \bmu_{\bbeta} - 2 \bbeta^T \bLambda_{\bbeta} \hat{\bbeta} +\hat{\bbeta}^T\bLambda_{\bbeta} \hat{\bbeta} \cr
  &\qquad +(\bbeta - \hat{\bbeta})^T \bLambda_{\bbeta} (\hat{\bbeta} - \bmu_{\bbeta}) + (\hat{\bbeta} - \bmu_{\bbeta})^T \bLambda_{\bbeta} (\bbeta - \bmu_{\bbeta})\cr
    &\quad= \by^T \by + (\bbeta - \hat{\bbeta})^T (\bX^T \bX + \bLambda_{\bbeta}) (\bbeta - \hat{\bbeta}) - \hat{\bbeta}^T(\bX^T \bX + \bLambda_{\bbeta}) \hat{\bbeta} \cr
    &\qquad -2 \bbeta^T \bLambda_{\bbeta} ( \hat{\bbeta} - \bmu_{\bbeta}) +\hat{\bbeta}^T\bLambda_{\bbeta} \hat{\bbeta} \cr
  &\qquad +2\bbeta^T \bLambda_{\bbeta} (\hat{\bbeta} - \bmu_{\bbeta}) - \hat{\bbeta}^T \bLambda_{\bbeta} (\hat{\bbeta} - \bmu_{\bbeta}) - \bmu_{\bbeta}^T \bLambda_{\bbeta} (\hat{\bbeta} - \bmu_{\bbeta}) \cr
  &\quad= \by^T \by + (\bbeta - \hat{\bbeta})^T (\bX^T \bX + \bLambda_{\bbeta}) (\bbeta - \hat{\bbeta}) - \hat{\bbeta}^T(\bX^T \bX + \bLambda_{\bbeta}) \hat{\bbeta} \cr
    &\qquad + \bmu_{\bbeta}^T \bLambda_{\bbeta} \bmu_{\bbeta},
 \end{align*}
 using that $\by^T \bX \bbeta = (\bX \bbeta)^T \by$ (because it is a scalar) and
 \begin{align*}
 \bbeta^T   (\bX^T \bX + \bLambda_{\bbeta}) \hat{\bbeta} =\bbeta^T (\bX^T \by + \bLambda_{\bbeta} \bmu_{\bbeta}).
 \end{align*}
 Therefore,
 \begin{align*}
  \pi(\bbeta, \xi \mid \by) &\propto \pi(\xi) \, \frac{1}{\xi^{n / 2}} \exp\left(-\frac{1}{2\xi}\left[\by^T \by - \hat{\bbeta}^T(\bX^T \bX + \bLambda_{\bbeta}) \hat{\bbeta} + \bmu_{\bbeta}^T \bLambda_{\bbeta} \bmu_{\bbeta}\right] \right) \cr
  &\quad \times \frac{1}{\xi^{p / 2}}\exp\left(-\frac{1}{2\xi}(\bbeta - \hat{\bbeta})^T (\bX^T \bX + \bLambda_{\bbeta}) (\bbeta - \hat{\bbeta}) \right).
 \end{align*}
From this, we can conclude that $\bbeta$ given $\xi$ has a normal distribution with a mean of $\hat{\bbeta}$ and a covariance matrix of $\xi (\bX^T \bX + \bLambda_{\bbeta})^{-1}$. Regarding $\xi$, we have that
\begin{align*}
 &\pi(\xi) \, \frac{1}{\xi^{n / 2}} \exp\left(-\frac{1}{2\xi}\left[\by^T \by - \hat{\bbeta}^T(\bX^T \bX + \bLambda_{\bbeta}) \hat{\bbeta} + \bmu_{\bbeta}^T \bLambda_{\bbeta} \bmu_{\bbeta}\right] \right) \cr
  &\quad\propto \frac{1}{\xi^{\frac{2 a + n}{2} + 1}} \exp\left(-\frac{1}{2\xi} \left[2b + \by^T \by - \hat{\bbeta}^T(\bX^T \bX + \bLambda_{\bbeta}) \hat{\bbeta} + \bmu_{\bbeta}^T \bLambda_{\bbeta} \bmu_{\bbeta}\right] \right),
\end{align*}
which allows to conclude that the posterior distribution of $\xi$ is an inverse-gamma with a shape parameter of $(2a + n) / 2$ and a scale parameter of
\[
 \frac{2b + \by^T \by - \hat{\bbeta}^T(\bX^T \bX + \bLambda_{\bbeta}) \hat{\bbeta} + \bmu_{\bbeta}^T \bLambda_{\bbeta} \bmu_{\bbeta}}{2}.
\]
\end{proof}

\subsection{Tukey’s biweight model}

In Sections \ref{sec:example1} and \ref{sec:example_reserve}, we use the following prior distribution for Tukey's biweight improper model: the distribution of $\bbeta$ is a normal with a mean of $\bmu_{\bbeta} \in \re^p$ and a (positive definite) covariance matrix of $\bSigma_{\bbeta}$. To estimate $\bbeta$, we sample from the generalized posterior distribution using Hamiltonian Monte Carlo (HMC). To run this algorithm, we need to evaluate the generalized posterior density up to a normalizing constant and to evaluate the gradient of the log density. We now write the posterior density (up to a normalizing constant), and next, the gradient of the log density:
\begin{align*}
 \pi(\bbeta \mid \by) &\propto \exp\left(-w \left(n \log\hat{\sigma} - \sum_{i=1}^n \log g\left(\frac{y_i - \bx_i^T \bbeta}{\hat{\sigma}}\right)\right)\right) \, \pi(\bbeta),
\end{align*}
where
\begin{align*}
-\log g(\varepsilon)=\left\{
\begin{array}{lcc}
                                                      1-(1-(\varepsilon/k)^2)^3  & \text{ if } & |\varepsilon|\leq k, \cr
                                                      1 & \text{ if } & |\varepsilon|>k.
\end{array}
\right.
\end{align*}

The log density is thus such that (if we forget about the normalizing constant):
\[
\log \pi(\bbeta \mid \by) =  \log \pi(\bbeta) -w \left(n \log\hat{\sigma} - \sum_{i=1}^n \log g\left(\frac{y_i - \bx_i^T \bbeta}{\hat{\sigma}}\right)\right).
\]
The gradient is such that:
\[
 \frac{\partial}{\partial \bbeta} \, \log \pi(\bbeta \mid \by) = - \bSigma_{\bbeta}^{-1} (\bbeta - \bmu_{\bbeta})  - \frac{w}{\hat{\sigma}} \sum_{i = 1}^n \frac{g'\left(\frac{y_i - \mathbf{x}_i^T \boldsymbol\beta}{\hat{\sigma}}\right)}{g\left(\frac{y_i - \mathbf{x}_i^T \boldsymbol\beta}{\hat{\sigma}}\right)} \, \mathbf{x}_i,
\]
where
\begin{align*}
\frac{g'(\varepsilon)}{g(\varepsilon)}=\left\{
\begin{array}{lcc}
            -\frac{6 \varepsilon}{k^2} \, \left(1 - \left(\frac{\varepsilon}{k}\right)^2\right)^2  & \text{ if } & |\varepsilon|\leq k, \cr
           0 & \text{ if } & |\varepsilon|>k.
\end{array}
\right.
\end{align*}

\section{Tukey's biweight scale estimator}\label{sec:asymptotic_scale}

In this section, we present the results of a simple numerical experiment to study the asymptotic behaviour of the estimator $\hat{\sigma}_{\text{TM}}$ that we propose to use in \autoref{sec:generalizedBayes}. The asymptotic framework is that where outliers move away from the bulk of the data. We consider the framework described in \autoref{sec:heavy-tailed}, meaning that some $y_i(\vartheta)$ go to $\pm\infty$ as $\vartheta \rightarrow \infty$ while the associated $\bx_i$ are kept fixed. We also consider the framework where, for some $i$, one component of $\bx_i(\vartheta)$, say $x_{ij}(\vartheta)$, go to $\pm\infty$ as $\vartheta \rightarrow \infty$ while the other components in $\bx_i(\vartheta)$ and $y_i$ are held fixed.

 In both cases, we first simulate a data set without outliers in the same way as described at the end of \autoref{sec:largesample} and then we gradually increase the values of some $y_i(\vartheta)$ or some $x_{ij}(\vartheta)$ (by increasing the value of $\vartheta$), obtaining a sequence of data sets. For each data set, we compute the resulting estimate $\hat{\sigma}_{\text{TM}}$, which we denote for the rest of the section by $\hat\sigma_\vartheta$ to highlight a dependence on $\vartheta$. We thus set $n = 20$, $p = 2$, $\bbeta_0 = (1, 1)^T$, $\sigma_0 = 1$ and $\mu_{\boldsymbol\bX}$ and $f_0$ both to the standard normal to obtain the data set without outliers.

\begin{figure}[ht]
 \centering\small
 $\begin{array}{cc}
   \includegraphics[width=0.5\textwidth]{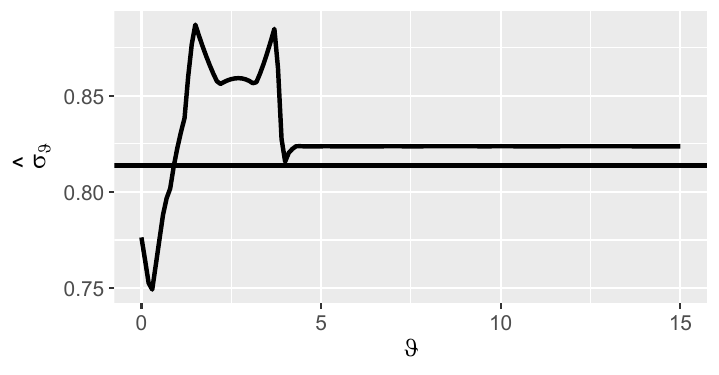} & \hspace{-5mm} \includegraphics[width=0.5\textwidth]{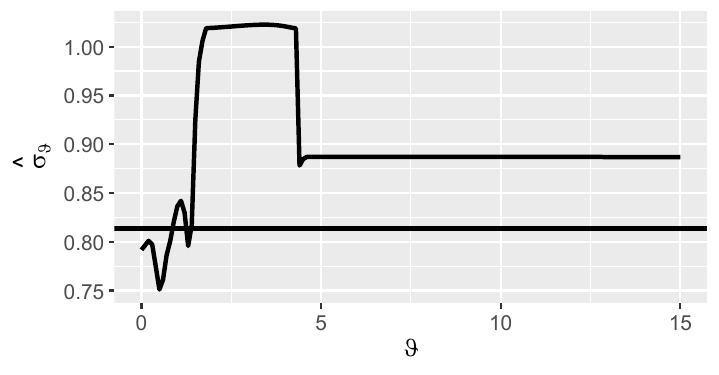} \cr
   \hspace{-0mm} \textbf{(a) 1 outlier} & \hspace{-5mm} \textbf{(b) 2 outliers}
  \end{array}$
  \vspace{-3mm}
\caption{$\hat\sigma_\vartheta$ as a function of $\vartheta$ in the case where some $y_i(\vartheta)$ go to $\infty$ and $k = 4.685$; in (a) there is one outlier, in (b) there are two outliers; the horizontal line represents the estimated value when the outliers are excluded.}\label{fig:sigma_4685}
\end{figure}

\begin{figure}[ht]
 \centering\small
 $\begin{array}{c}
   \includegraphics[width=0.5\textwidth]{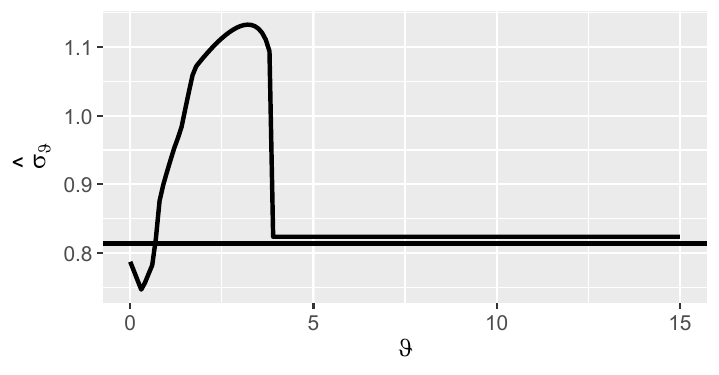}  \cr
   \hspace{-0mm} \textbf{(a) 1 outlier}
  \end{array}$
  \vspace{-3mm}
\caption{$\hat\sigma_\vartheta$ as a function of $\vartheta$ in the case where some $x_{ij}(\vartheta)$ go to $\infty$ for one outlier and $k = 4.685$; the horizontal line represents the estimated value when the outliers are excluded.}\label{fig:sigma_4685x}
\end{figure}

 In \autoref{fig:sigma_4685}, we present the results in the case where some $y_i(\vartheta)$ go to $\infty$ and $k = 4.685$. When there is one outlier, $\hat\sigma_\vartheta$ is larger than the value without the outlier by 1\% when $\vartheta$ is large. When there are two outliers, it is larger by 9\%. When $k = 4.685$, the breakdown point evaluated in this example is $3 / 20 = 15\%$; in the presence of 4 outliers, $\hat\sigma_\vartheta$ increases with $\vartheta$. In \autoref{fig:sigma_4685x}, we present the results in the case where some $x_{ij}(\vartheta)$ go to $\infty$ and $k = 4.685$. We only present the results in the case where there is one outlier because, when there is more than one outlier, the convergence only happens when $\vartheta$ is extremely large. When there is one outlier, $\hat\sigma_\vartheta$ is again larger than the value without the outlier by 1\% when $\vartheta$ is large.

\normalsize

\normalsize

\begin{figure}[ht]
 \centering\small
 $\begin{array}{cc}
   \includegraphics[width=0.5\textwidth]{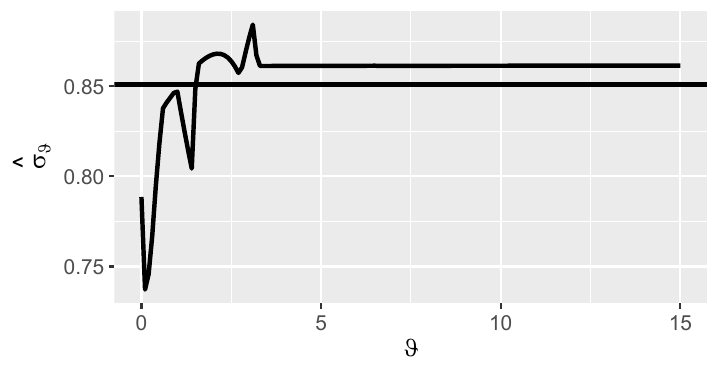} & \hspace{-5mm} \includegraphics[width=0.5\textwidth]{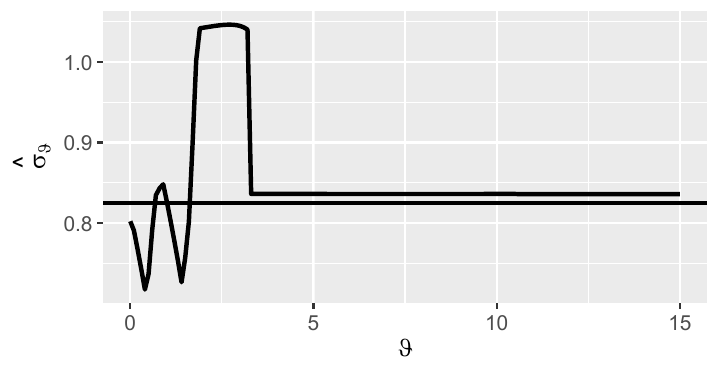} \cr
   \hspace{-0mm} \textbf{(a) 1 outlier} & \hspace{-5mm} \textbf{(b) 2 outliers}
  \end{array}$
  \vspace{-3mm}
\caption{$\hat\sigma_\vartheta$ as a function of $\vartheta$ in the case where some $y_i(\vartheta)$ go to $\infty$ and $k = 3$; in (a) there is one outlier, in (b) there are two outliers; the horizontal line represents the estimated value when the outliers are excluded.}\label{fig:sigma_3}
\end{figure}

\begin{figure}[ht]
 \centering\small
 $\begin{array}{cc}
   \includegraphics[width=0.5\textwidth]{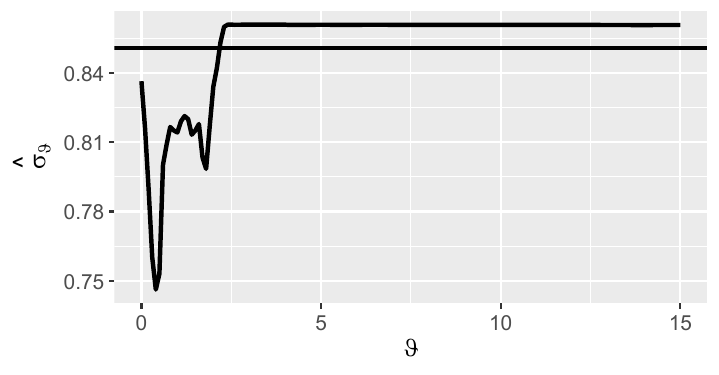} & \hspace{-5mm} \includegraphics[width=0.5\textwidth]{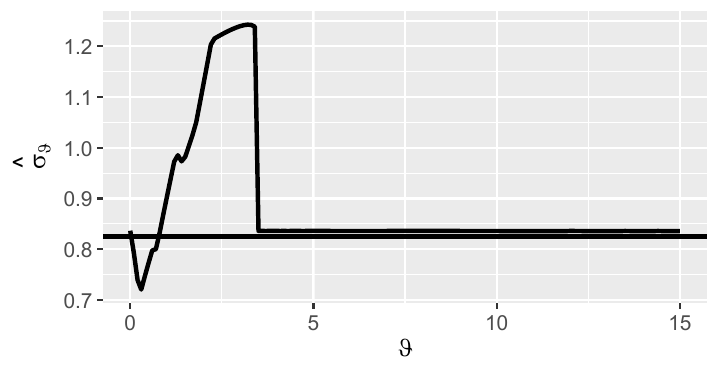} \cr
   \hspace{-0mm} \textbf{(a) 1 outlier} & \hspace{-5mm} \textbf{(b) 2 outliers}
  \end{array}$
  \vspace{-3mm}
\caption{$\hat\sigma_\vartheta$ as a function of $\vartheta$ in the case where some $x_{ij}(\vartheta)$ go to $\infty$ and $k = 3$; in (a) there is one outlier, in (b) there are two outliers; the horizontal line represents the estimated value when the outliers are excluded.}\label{fig:sigma_3x}
\end{figure}

In \autoref{fig:sigma_3}, we present the results in the case where some $y_i(\vartheta)$ go to $\infty$ and $k = 3$. In the presence of one and two outliers, $\hat\sigma_\vartheta$ is larger by 1\%. In this case, the breakdown point is $4/20 = 20\%$. In \autoref{fig:sigma_3x}, we present the results in the case where some $x_{ij}(\vartheta)$ go to $\infty$ and $k = 3$. In the presence of one and two outliers, $\hat\sigma_\vartheta$ is again larger by 1\%.

\normalsize

\normalsize

\section{Justification of the definition of $\pi_n$ in \autoref{sec:largesample}}\label{sec:justification_pi_n}

In this section, we provide more details justifying that the generalized posterior distribution studied in \autoref{sec:largesample} is that defined in \eqref{eq:postTukey}, even if each $\bx_i$ is considered a realization of a random vector $\bX_i$. For the explanation, let us first consider a proper Bayesian linear regression model as in \autoref{sec:heavy-tailed} with $f$ the PDF of the error term in this model, but with $\bx_i$ a realization of $\bX_i$ and $\sigma = \hat\sigma_{\text{TM}}$, the robust estimator that we propose to use in \autoref{sec:generalizedBayes}. The posterior distribution of $\bbeta$ in this case is a conditional distribution and we consider here that it is given $Y_1 = y_1, \ldots, Y_n = y_n$ and $\bX_1 = \bx_1, \ldots, \bX_n = \bx_n$. Assuming that $\bX_1, \ldots, \bX_n, \varepsilon_1, \ldots, \varepsilon_n$ and $\bbeta$ are all independent in the Bayesian model, the posterior distribution does not in fact depend on the assumed distribution of $\bX_i$; we can thus assume in the Bayesian model that $\bX_i \sim \mu_{\bX}$. Indeed, the posterior density is such that
\begin{align*}
  \pi(\bbeta \mid \by, \bx_1, \ldots, \bx_n) &:= \frac{\pi(\bbeta) \prod_{i=1}^n \frac{1}{\hat\sigma_{\text{TM}}} f\left(\frac{y_i - \bx_i^T \bbeta}{\hat\sigma_{\text{TM}}}\right) \, f_{\bX}(\bx_i)}{\int_{\re^p} \pi(\bbeta) \prod_{i=1}^n \frac{1}{\hat\sigma_{\text{TM}}} f\left(\frac{y_i - \bx_i^T \bbeta}{\hat\sigma_{\text{TM}}}\right) \, f_{\bX}(\bx_i) \, \d\bbeta} \cr
   &= \frac{\pi(\bbeta) \prod_{i=1}^n \frac{1}{\hat\sigma_{\text{TM}}} f\left(\frac{y_i - \bx_i^T \bbeta}{\hat\sigma_{\text{TM}}}\right) }{\int_{\re^p} \pi(\bbeta) \prod_{i=1}^n \frac{1}{\hat\sigma_{\text{TM}}} f\left(\frac{y_i - \bx_i^T \bbeta}{\hat\sigma_{\text{TM}}}\right)  \, \d\bbeta},
\end{align*}
assuming that the denominator is finite and abusing notation by writing $f_{\bX}(\bx_i)$ given that $f_{\bX}$ is the PDF of all components of $\bX_i$ except the first one. This posterior distribution is the \emph{same} as when each $\bx_i$ is not considered a realization of a random vector. It is not possible to derive the generalized posterior distribution under Tukey's biweight improper model as above (because it is not a conditional distribution). However, to be consistent with the case above where the linear regression model is proper, we should define the generalized posterior distribution under Tukey's biweight improper model as above with $f = g$ given in \eqref{eqn:biweight}, corresponding to the definition in \eqref{eq:postTukey}, up to the tempering parameter $w > 0$ which is added for calibration purposes.

\end{document}